\documentclass[%
 reprint,
 amsmath,amssymb,
 aps,
]{revtex4-2}

\usepackage{graphicx}
\usepackage{dcolumn}
\usepackage{bm}

\usepackage{enumitem}

\setlist[itemize]{leftmargin=*}

\usepackage{amsmath, amsthm, amssymb, amsfonts,mathrsfs}

\usepackage[linktocpage=true]{hyperref}
\hypersetup{
	colorlinks,
	citecolor=blue,
	filecolor=black,
	linkcolor=blue,
	urlcolor=blue
}
\usepackage{url}
  
\usepackage{physics}

\usepackage{titlesec}
\titlespacing*{\subsection}{0pt}{1.5ex plus 1ex minus .2ex}{1ex plus .2ex}

\begin{document}

\preprint{APS/123-QED}


\title{Scalar-Field Wave Dynamics and Quasinormal Modes of the Teo Rotating Wormhole}

\author{Ramesh Radhakrishnan}
\email{ramesh\_radhakrishna1@baylor.edu}
\affiliation{Department of Physics, Baylor University, Waco, Texas 76798, USA}
\affiliation{Early Universe, Cosmology and Strings (EUCOS) Group, 
Center for Astrophysics, Space Physics and Engineering Research (CASPER), 
Baylor University, Waco, Texas 76798, USA}

\author{Delaram Mirfendereski}
\email{delaram.mirfendereski@utrgv.edu}
\affiliation{Department of Physics and Astronomy, 
The University of Texas Rio Grande Valley, USA}
\affiliation{Early Universe, Cosmology and Strings (EUCOS) Group, 
Center for Astrophysics, Space Physics and Engineering Research (CASPER), 
Baylor University, Waco, Texas 76798, USA}

\author{Eric W.~Davis}
\email{ewdavis@albany.edu}
\affiliation{Department of Physics, 
State University of New York at Albany, Albany, New York 12222, USA}
\affiliation{Early Universe, Cosmology and Strings (EUCOS) Group, 
Center for Astrophysics, Space Physics and Engineering Research (CASPER), 
Baylor University, Waco, Texas 76798, USA}

\author{Claudio Cremaschini}
\email{claudio.cremaschini@physics.slu.cz}
\affiliation{Institute of Physics and Research Center for Theoretical Physics and Astrophysics, 
Faculty of Philosophy and Science, Silesian University in Opava, 
Bezru{\v{c}}ovo n{\'a}m.\ 13, CZ-74601 Opava, Czech Republic}

\author{Gerald~B.~Cleaver}
\email{gerald_cleaver@baylor.edu}
\affiliation{Department of Physics, Baylor University, Waco, Texas 76798, USA}
\affiliation{Early Universe, Cosmology and Strings (EUCOS) Group, 
Center for Astrophysics, Space Physics and Engineering Research (CASPER), 
Baylor University, Waco, Texas 76798, USA}

\date{\today}

\begin{abstract}
In this work, we investigate scalar field perturbations of the rotating Teo wormhole spacetime. We also compute the quasinormal mode (QNM) spectrum using first order WKB approximation for a geometry devoid of horizons. After separation of variables, we obtain a Schr\"odinger type radial equation and a smooth barrier potential which is shaped by the localized frame-dragging effects of the wormhole throat. This barrier potential provides damped oscillatory modes for the range of spins that were examined. 

The QNM spectrum thus obtained, shows a coherent and monotonic dependence on rotation. As the spin increases, both the oscillating frequency of the scalar wave $\Re(\omega)$ and its damping rate $-\Im(\omega)$ decrease, which indicates progressively longer lived modes in the absence of absorption due to a horizon. We have also verified the correspondence $\omega_{\mathrm{QNM}} \simeq m\,\Omega_{\mathrm{ph}} - i(n+\tfrac{1}{2})\lambda$ in the Eikonal limit, by obtaining the radius of the photon ring, its orbital frequency, and the Lyaponov exponent. 

Next, we compared the Teo wormhole QNM with that of the Kerr black hole QNM to find that the Kerr QNM is dictated by absorption at the horizon and they also exhibit symmetric prograde/retrograde mode splitting, whereas the Teo wormhole QNM shows a stronger, and spatially confined response to spin. The Teo wormhole also exhibit partial reflection at the throat and a very distinct one-sided $m=\pm2$ mode splitting which rapidly saturates as the spin increases. Additionally, the rotating Teo wormhole allows an ergoregion with the possibility of frequency kinematics compatible with superradiance. However, due to the absence of an event horizon or a dissipative boundary, there is no evidence of classical superradiant amplification that was seen in Kerr. 

The results we obtained clearly demonstrates how rotation and boundary conditions jointly shape wave propagation in horizonless compact objects. They also provide certain characteristic spectral signatures that can be used to distinguish rotating wormhole spacetimes from rotating black hole spacetimes. 
\end{abstract}

\maketitle


\setcounter{tocdepth}{2} 
\tableofcontents
\clearpage

\section{\label{sec:intro}Introduction}

The well known Morris-Thorne wormhole geometry was modified by Teo \cite{Teo1998} to include angular momentum, resulting in the rotating Teo wormhole spacetime, which is a useful model. It extends the construction by Morris and Thorne \cite{MorrisThorne1988} to a stationary, axisymmetric, and traversable rotating wormhole spacetime geometry. The Teo wormhole has a smooth throat, and no horizons. It's rotation parameter introduces key features such as frame-dragging, mode-splitting, and ergoregion with possibility for superradiance compatible kinematics. 

Since most compact astrophysical objects from stars and planets to galaxies exhibit rotation, the investigation of a rotating wormhole spacetime is natural and motivated both by its physics and observations of rotating astrophysical objects. In his classical paper \cite{Teo1998}, Teo also showed the possibility of an ergoregion that surrounds the throat and that frame-dragging is a defining feature of the ergoregion. The Teo wormhole also violates the Null Energy Condition (NEC), in at least some regions near the throat, (for example, \ $T_{\hat{t}\hat{t}}+T_{\hat{r}\hat{r}}<0$). Violation of the NEC is a requirement for traversable wormholes. 

Both Visser \cite{Visser1989}, and Teo \cite{Teo1998} showed that the NEC violation can be restricted to limited regions instead of being distributed uniformly around the throat. So, rotation partially loclizes this exotic matter requirement/violation of NEC and the spatial distribution of negative energy regions depend on rotation profile choices. For suitable model parameters, it was shown by Teo, that there exists a class of both timelike and null geodesics that traverse the wormhole throat without entering the regions that violate the NEC. Visser had also observed earlier that there are wormhole geometries that can be built, in which a traveller follows a timelike geodesic that avoids zones of exotic matter. There were investigations conducted later that also showed quantitatively, how the total amount of energy-condition violation can be made very small if one can carefully engineer wormhole geometries. Thus, although for global traversability, negative energy regions or the need for exotic matter cannot be avoided, rotation and frame-dragging allow a partial redistribution of exoticity locally which permits certain specfic trajectories that can experience an effective ``normal" spacetime. Therefore, the Teo rotating wormhole metric acts as a theoretical laboratory in which we can investigate and understand wave propagation, traversability, and stability in rotating wormhole backgrounds. 

Gravitational wave detectors such as LIGO, Virgo, and KAGRA have opened up a plethora of investigations and opportunities to analyze the strong gravity regime \cite{Abbott2016,Abbott2021GWTC3,Abbott2019TestsGR}. The gravitational signals received by these detectors include the classic ``ringdown" signatures observed during the mergers of massive and supermassive compact objects such as black holes and neutron stars. The ringdown spectrum corresponds to the exponentially damped oscillations of hte merger remnant, when it is settling down into a stable configuration. 

We know that the observed ringdown spectra see in gravitational waves are consistent with the predictions of General Relativity (GR) about the existence of relativistic black holes. However, there are solutions to Einstein's equations that are other exotic compact objects (ECOs) such as wormholes. Current researchers have great interest in checking if these ECOs including wormholes, would have a QNM spectrum that is similar to that of black holes, or if they are different, in what ways do they differ \cite{Cardoso2016,Abedi2017,Conklin2023}. 

There has also been search efforts fro late-time ``echoes", that are secondary gravitational pulses created due to partial reflections of the initial ringdown signal within the compact objects' interior (for example, reflection from the wormhole throat) or near-horizon structure. These effects including deviation from rotating Kerr black hole ringdown spectra have not given us any conclusive results, except for statistically marginal results \cite{Westerweck2018,Uchikata2023}. This motivates us to continue work on theoretical modeling of wave dynamics and scattering in such ECOs that may provide distinct signatures that are deviations from that of black holes. 

A deeper understanding of the QNM spectrum of wormholes is critical in order to predict their properties such as characteristic frequencies, damping times, and potential echo delays that could help us distinguish clearly between black hole and wormhole QNM spectra. 

QNMs are complex oscillating frequencies that are usually seen as the response of black hole spacetimes to perturbations such as a scalar field perturbation \cite{Cardoso_2001,Berti-2009}, and the study of QNMs is of great importance in the field of gravitational wave research \cite{FerrariMashhoon1984,PhysRevD.62.024027}. The characteristic QNM wave form \cite{Hawking:1975vcx,Hawking:1974rv} carries information about the reacton of the spacetimes such as black holes and wormholes, and their properties such as mass, charge and angular momentum. The stability of the Teo rotating wormhole spacetime can be investigated \cite{PhysRevD.98.084011,guo2021scalarized} using the imaginary part of the complex QNM frequency, which is tied to the damping scale, and since it is hard to find an exact analytical solution for the QNM eigen frequencies, several approximation methods have been used in the past, such as the WKB approximation \cite{SchutzWill1985}. Others have also used higher-order corrections to WKB to increase the accuracy of the solution \cite{IyerWill1987,PhysRevD.41.374}. Studies similar to the one in this paper for axially symmetric Teo wormhole, have been conducted for spherically symmetric compact objects \cite{Franzin_2024}.

In this paper, we analyze scalar-field dynamical interaction in the form of perturbations on the axisymmetric Teo rotating wormhole background. This is a tractable first step toward a full tensorial perturbation study. we use a minimally coupled scalar field which serves as a test probe and captures the
essential physics of wave propagation and potential barriers in curved spacetimes. We avoid the technical complexities of metric perturbations
\cite{Konoplya_2011,PhysRevD.98.044035}, which can be a great subsequent study. 
By analyzing the Klein-Gordon equation in the Teo spacetime geometry, we derive an effective potential $V_{\mathrm{eff}}(r)$ governing the radial propagation of scalar waves.
The multipole number $\ell$ usually labels the angular structure of scalar perturbations and enters the effective potential as a parameter that controls
the barrier height. The azimuthal number $m$ couples to the wormhole rotation and introduces a frame-dragging dependence in $V_{\mathrm{eff}}$. In our current calculations, we explore representative $m$-values. The effective potential includes the rotational $m$-dependence, and we observe mode splitting, however, the full pattern of $m$-splitting across all azimuthal numbers is not taken into account. 

We perform an exploratory computation of the associated QNMs using a first-order Wentzel--Kramers--Brillouin (WKB) approximation
\cite{SchutzWill1985,Konoplya2003,Konoplya-Zinhailo}, which relies on the peak of $V_{\mathrm{eff}}(r)$ and its derivatives. The analysis of $V_{\mathrm{eff}}$ provides direct insight into the scattering
properties of the wormhole throat, including possible single- or double-peak features that can create long-lived modes, that are also called trapped modes or quasi-bound states. These trapped modes correspond to oscillations that decay very slowly due to partial
confinement of the scalar field between potential barriers or wells
\cite{Cardoso2009,Churilova_2020,Cardoso_2016}.
Such long-lived modes are characterized by a small imaginary part of the complex QNM frequency. This implies slow damping, and in some parameter regimes they may produce late-time echoes in the waveform, though echoes are not guaranteed.

Our goal here is to map the qualitative dependence of the fundamental QNM frequencies on the wormhole rotation parameter ($a$), throat size ($r=b_0$), azimuthal number ($m$), and overtone number ($n$). We also delineate the parameter regimes where the first-order WKB approximation provides reliable predictions. We then track $m$-splitting, ergoregion formation, and superradiant-compatible kinematics. Other rotational effects may also be observed.

\section{Teo Wormhole Geometry and Perturbation Setup}
\subsection{\label{sec:scat}Klein--Gordon equation for the Teo wormhole}

We begin with the Klein--Gordon (KG) equation for a massless scalar field
\begin{equation}
    \Box\Phi = \frac{1}{\sqrt{-g}}\,
    \partial_{\mu}\!\Big(\sqrt{-g}\, g^{\mu\nu} \partial_{\nu}\Phi\Big)=0,
    \label{eq:KG-master}
\end{equation}
propagating on the rotating Teo wormhole geometry.  Throughout this work we adopt the equatorial-plane specialization of the Teo metric, which retains all rotational effects while simplifying the algebraic structure of the KG operator.

\subsubsection*{Metric functions and equatorial-plane reduction}
The full Teo line element is
\begin{multline}\label{eq:Teo_line_element}
ds^2=-N^2(r,\theta)\, dt^2
+ \frac{dr^2}{1 - {b_0}/{r^2}} 
+ r^2 K^2(r,\theta)\, d\theta^2 \\
+ r^2 K^2(r,\theta)\sin^2\theta\,
  \bigl(d\phi-\Omega(r)\,dt\bigr)^2,
\end{multline}
where 
\begin{equation}\label{eq:frame_dragging}
\Omega(r)=2a/r^3
\end{equation}
generates frame dragging. For analytic tractability and because the qualitative rotational physics is unaffected, we evaluate the metric at $\theta=\pi/2$ with $N(r,\theta)=K(r,\theta)=1$. For equatorial trajectories, we further restrict to $d\theta=0$.

The nonvanishing metric components then reduce to
\begin{align}
    g_{tt} &= -1 + r^{2}\Omega^2(r), \\
    g_{t\phi} &= -r^{2}\Omega(r), \\
    g_{\phi\phi} &= r^{2},\\
    g_{rr} &= \left(1 - \frac{b_0}{r^2}\right)^{-1},\\
    g_{\theta\theta} &=r^{2},
\end{align}
with inverse components
\begin{align}
    g^{tt} &= -1, \qquad 
    g^{t\phi}=-\Omega(r), \qquad
    g^{\phi\phi}=\frac{1-r^{2}\Omega^2(r)}{r^{2}},\\[4pt]
    g^{rr} &= 1-\frac{b_0}{r^2}, \qquad
    g^{\theta\theta}=\frac{1}{r^{2}}.
\end{align}
The determinant simplifies to
\begin{equation}
    \sqrt{-g}=\,r^{2}\Bigl(1 - \frac{b_0}{r^2}\Bigr)^{-1/2}.
\end{equation}

\subsubsection*{Mode decomposition}
Stationarity and axial symmetry allow the decomposition
\begin{equation}
\Phi(t,r,\theta,\phi)
   = e^{-i\omega t} e^{im\phi}\, \Psi(r,\theta),
\end{equation}
so that the time--angular sector of the KG operator becomes
\begin{equation}
\omega^2 g^{tt}
  - 2m\omega g^{t\phi}
  + m^{2}g^{\phi\phi}
= -\bigl(\omega-m\Omega(r)\bigr)^2
  + \frac{m^{2}}{r^{2}}.
\label{eq:time-ang}
\end{equation}

We are interested in scalar configurations that remain confined to the
equatorial plane, $\partial_\theta\Phi=0$, so that all $\theta$-derivatives in
\eqref{eq:KG-master} drop out.  The KG equation then takes the form
\begin{equation}
\Bigl[-(\omega-m\Omega)^2+\frac{m^{2}}{r^{2}}\Bigr]\Phi
+ \frac{1}{\sqrt{-g}}
  \partial_r\!\Bigl(\sqrt{-g}\,g^{rr}\Phi'\Bigr)
= 0.
\label{eq:KG-equatorial}
\end{equation}

\subsubsection*{Radial operator and tortoise coordinate}
Using the explicit form of $\sqrt{-g}\,g^{rr}$,
\[
\sqrt{-g}\,g^{rr}
= r^{2}\sqrt{1-\frac{b_0}{r^{2}}},
\]
the radial operator becomes
\begin{equation}
\frac{1}{\sqrt{-g}}\partial_r\!\bigl(\sqrt{-g}\,g^{rr}\Phi'\bigr)
=
\frac{\sqrt{1-b_0/r^{2}}}{r^{2}}
\frac{d}{dr}\!\left(r^{2}\sqrt{1-\frac{b_0}{r^{2}}}\,\Phi'\right).
\label{eq:radial-operator}
\end{equation}
It is convenient to introduce the tortoise coordinate
\begin{equation}
\frac{dr_*}{dr}
= \frac{1}{\sqrt{1 - b_0/r^{2}}},
\label{eq:tortoise}
\end{equation}
which regularizes the throat at $r^2=b_0$.  In terms of $r_*$, the radial operator becomes
\begin{equation}
\frac{d^{2}\Phi}{dr_*^{2}}
+ P(r)\frac{d\Phi}{dr_*},
\qquad
P(r)=\frac{2}{r}\sqrt{1-\frac{b_0}{r^{2}}}.
\label{eq:operator-rstar}
\end{equation}
Thus, even in the tortoise coordinate a first-derivative term remains, so an
additional field redefinition is required before a Schr\"odinger-type equation
can emerge.

\subsubsection*{Liouville transformation and Schr\"{o}dinger form}

To eliminate the first-derivative term we introduce the standard Liouville
transformation
\begin{equation}
\Phi(r_*) = A(r)\,\Psi(r_*),
\qquad 
\frac{A'}{A} = -\frac{P}{2},
\label{eq:Liouville}
\end{equation}
where a prime denotes differentiation with respect to $r_*$.  Substituting
\eqref{eq:Liouville} into \eqref{eq:KG-equatorial} yields the
Schr\"{o}dinger-type master equation
\begin{equation}
\frac{d^{2}\Psi}{dr_*^{2}}
+ \Bigl[\omega^{2}-V_{\rm eff}(r)\Bigr]\Psi = 0.
\label{eq:Schrod-final}
\end{equation}
The effective potential is
\begin{equation}
V_{\rm eff}(r)
= (\omega-m\Omega)^2 - \frac{m^{2}}{r^{2}}
+ \frac{P'}{2} + \frac{P^{2}}{4},
\label{eq:Veff-correct}
\end{equation}
where $P(r)$ is given by \eqref{eq:operator-rstar}.  The last two terms encode
the geometric contributions arising from the radial measure and are essential for
placing the KG equation into Sturm--Liouville form.  These terms must be retained
(even when small) for any WKB analysis based on the Schr\"odinger-type equation
\eqref{eq:Schrod-final}.

\subsubsection*{Relation to the co-rotating KG potential}
If one temporarily omits the geometric corrections $P'/2$ and $P^{2}/4$
—as is appropriate only in the geometric--optics limit—the potential reduces to the algebraic Klein--Gordon form
\begin{equation}
V_{\rm eff}(r)
=(\omega-m\Omega)^2-\frac{m^{2}}{r^{2}},
\end{equation}
which is the expression used for photon-ring analysis and for establishing the geometric correspondence between quasinormal modes and unstable circular null orbits. In the eikonal (geometric--optics) limit, we therefore identify this quantity as the effective potential $V_{\rm eff}^{\rm (eikonal)}$. In contrast, the Schr\"odinger potential \eqref{eq:Veff-correct} must be used for WKB quasinormal-mode calculations.

\section{Effective Potential and Circular Photon Orbits}
For null geodesics confined to the equatorial plane of the rotating Teo wormhole spacetime, an effective radial potential governs the motion of photons and the formation of photon rings.  The frame-dragging angular velocity is given by \eqref{eq:frame_dragging}.

From the Klein--Gordon reduction in the equatorial plane, the radial part of the separated wave equation naturally suggests an \emph{eikonal} effective potential of the form
\begin{equation}
V_{\text{eff}}(r)
   = \big(\omega - m\,\Omega(r)\big)^{2}
     - \frac{m^{2}}{r^{2}},
\label{eq:Veff_eikonal_def}
\end{equation}
where $\omega$ and $m$ label the Fourier mode of the scalar field.  For the canonical Teo wormhole with $\Omega(r)=2a/r^{3}$, this becomes
\begin{equation}
V_{\text{eff}}(r) = -\frac{m^{2}}{r^{2}} + \frac{(\omega r^{3} - 2 a m)^{2}}{r^{6}},
\label{eq:Veff_expanded}
\end{equation}
whose first derivative is
\begin{equation}
V'_{\text{eff}}(r)
= \frac{12 a m \omega}{r^{4}}
 - \frac{24 a^{2} m^{2}}{r^{7}}
 + \frac{2 m^{2}}{r^{3}},
\end{equation}
and the second derivative, governing local stability, is
\begin{equation}
V''_{\text{eff}}(r)
= -\frac{48 a m \omega}{r^{5}}
 + \frac{168 a^{2} m^{2}}{r^{8}}
 - \frac{6 m^{2}}{r^{4}}.
\end{equation}
The sign of \(V''_{\text{eff}}(r)\) at an extremum distinguishes unstable (\(V''<0\)) from stable (\(V''>0\)) circular configurations in the eikonal wave picture.

Here, \( \omega \) denotes the wave-mode frequency associated with the temporal oscillation of the perturbation, while \( \Omega(r) \) represents the frame-dragging angular velocity of the rotating wormhole geometry.  In the geometric-optics (high-frequency) limit, the dispersion relation implied by \eqref{eq:Veff_eikonal_def} coincides with the Hamilton--Jacobi equation for null geodesics, so the same radial structure governs both the eikonal KG modes and the photon trajectories.

\subsection{Photon Rings and Stability Analysis}
In the context of the Teo rotating wormhole~\cite{Teo1998}, photon rings correspond to unstable circular null geodesics that form the boundary of the observable shadow or bright emission ring seen by distant observers~\cite{Perlick2022}. These orbits occur at the radius \(r = r_{\text{ph}}\) where photons can temporarily orbit the wormhole throat under the balance of centrifugal repulsion and frame-dragging effects.

In this work we treat the photon ring in two complementary ways. On the one hand, we obtain $r_{\text{ph}}(a,b)$ from the \emph{null geodesic equations} by solving the circular-orbit conditions $F(r,b)=0$ and $\partial_r F(r,b)=0$, where $b=L/E$ is the impact parameter. Here $F(r,b)$ denotes the radial null--geodesic potential obtained from the Hamiltonian constraint $g^{\mu\nu}p_\mu p_\nu=0$, with conserved energy $E\equiv -p_t$ and angular momentum $L\equiv p_\phi$ arising from stationarity and axial symmetry, respectively. Circular photon orbits correspond to extrema of $F(r,b)$.

On the other hand, the eikonal Klein--Gordon potential \eqref{eq:Veff_eikonal_def} provides a convenient wave-based proxy: in the geometric-optics limit, the phase $\Phi\sim e^{iS/\hbar}$ obeys the Hamilton--Jacobi equation $g^{\mu\nu}\partial_\mu S\,\partial_\nu S = 0$, which is equivalent to the null geodesic condition~\cite{FerrariMashhoon1984,Cardoso2009}. Consequently, the extrema of $V_{\text{eff}}(r)$ track the circular null orbits, and its curvature at the peak encodes the same instability information as the Lyapunov exponent computed from geodesic deviation~\cite{FerrariMashhoon1984,Mashhoon1985,Cardoso2009}.

The photon ring radius is determined by the geodesic conditions
\begin{equation}
F(r_{\text{ph}},b)=0,
\qquad
\partial_r F(r,b)\big|_{r=r_{\text{ph}}}=0,
\end{equation}
which identify an unstable circular null orbit. In the eikonal picture, one may equivalently locate a nearby radius by solving
\begin{equation}
\frac{dV_{\text{eff}}}{dr}\Big|_{r=r_{\text{peak}}} = 0,
\qquad
\frac{d^{2}V_{\text{eff}}}{dr^{2}}\Big|_{r=r_{\text{peak}}} < 0,
\end{equation}
for fixed $(\omega,m)$, with $r_{\text{peak}}$ lying close to $r_{\text{ph}}$ in the high-frequency limit.

The quantity $\Omega_{\text{ph}} = (d\phi/dt)_{r_{\text{ph}}}$ represents the orbital frequency of photons at the unstable circular orbit, corresponding to the frame-dragging rate measured by a distant observer. In the eikonal regime, this frequency is related to the real part of the quasinormal frequency via
\[
\omega_{R}/m \simeq \Omega_{\text{ph}},
\]
providing a direct bridge between photon-ring properties and the wave spectrum.

To quantify the orbit's instability, we compute the Lyapunov exponent $\lambda$~\cite{Cardoso2009}:
\begin{equation}
\lambda
= \sqrt{-\frac{V''_{\text{geo}}(r_{\text{ph}})}{2 \dot t^{\,2}}},
\end{equation}
where $V_{\text{geo}}(r)$ denotes the \emph{geodesic} radial potential and $\dot t = dt/d\tau$ is obtained from the first integrals. For equatorial null motion in the Teo metric,
\begin{equation}
ds^{2} = -N^{2}(r)\,dt^{2}
+ \frac{dr^{2}}{1-b_{0}/r^{2}}
+ r^{2}[d\phi - \Omega(r)\,dt]^{2},
\end{equation}
the conserved quantities
\[
E=-g_{tt}\dot t - g_{t\phi}\dot\phi,
\qquad
L=g_{t\phi}\dot t+g_{\phi\phi}\dot\phi
\]
lead to
\begin{equation}
\dot t = \frac{E g_{\phi\phi} + L g_{t\phi}}{g_{t\phi}^{2} - g_{tt}g_{\phi\phi}}.
\end{equation}
At the circular photon orbit this simplifies to
\begin{equation}
\dot t = \frac{E}{N^{2}(r)},
\end{equation}
so that
\begin{equation}
\lambda_{t}
= \frac{N^{2}(r_{\text{ph}})}{E}\,
  \sqrt{-\frac{V''_{\text{geo}}(r_{\text{ph}})}{2}}.
\end{equation}
Here the subscript $t$ indicates that the Lyapunov exponent is defined with respect to the coordinate time $t$ measured by a distant observer, rather than the affine parameter along the null geodesic. With the standard normalization \(E=1\) and $N=1$ for the Teo wormhole,
\begin{equation}
\boxed{
\lambda_{t}
= \sqrt{-\frac{V''_{\text{geo}}(r_{\text{ph}})}{2}}
}.
\end{equation}
In the eikonal limit the curvature of $V_{\text{geo}}$ at $r_{\text{ph}}$ coincides with that of the KG-derived potential $V_{\text{eff}}$ to leading order, so one may interchange them for practical computations of $\lambda$.

A larger $\lambda$ indicates stronger local instability of the orbit. These quantities are plotted in Fig.~\ref{fig:rph_vs_a}. The photon ring consistently lies outside the ergosurface radius $r_{\text{ergo}}=\sqrt{2a}$, mirroring the geometric ordering in Kerr.

\subsection{On the Unstable Nature of the Photon Ring}
The photon ring in the rotating Teo wormhole arises from the \emph{unstable} branch of the effective potential, characterized by $V_{\text{geo}}''(r_{\text{ph}})<0$. At this location, photons occupy circular null geodesics at the crest of the potential barrier: even infinitesimal radial perturbations cause them to escape outward or fall inward toward the throat. This instability defines the sharp optical edge of the photon ring in ray-tracing simulations~\cite{Perlick2022}. The Lyapunov exponent $\lambda$ quantifies the divergence of nearby photon trajectories and directly determines the imaginary part of the corresponding quasinormal modes~\cite{Cardoso2009}. This mirrors the situation in black holes such as Kerr~\cite{Johannsen2013}. Thus, the unstable branch of the radial potential encodes both optical and perturbative signatures of the Teo wormhole spacetime.

\subsection{Connection to Quasinormal Modes}
In the eikonal limit, where the perturbation wavelength is much smaller than the curvature radius, the dynamics of wave propagation is governed by the same barrier that controls the unstable circular null orbits. Consequently, the photon-ring parameters $(r_{\text{ph}},\Omega_{\text{ph}},\lambda)$ extracted from the null geodesics (or equivalently from $V_{\text{eff}}$ in the geometric-optics limit) encode the leading-order behavior of quasinormal modes~\cite{FerrariMashhoon1984,Cardoso2009,Stefanov2010}. The real and imaginary parts of the QNM frequency follow
\begin{equation}
\boxed{
\omega_{\text{QNM}}
\simeq m\,\Omega_{\text{ph}}
 - i\left(n + \tfrac{1}{2}\right)\lambda
}.
\end{equation}
This correspondence unifies geometric optics and wave dynamics in the Teo wormhole spacetime and provides a useful benchmark for the full WKB analysis based on the Schr\"odinger-form potential.

\subsection{Variation of the Klein--Gordon Effective Potential with the Rotation Parameter}
\label{subsec:Veff_a_variation}

\begin{figure}[t]
    \centering
    \includegraphics[width=\columnwidth]{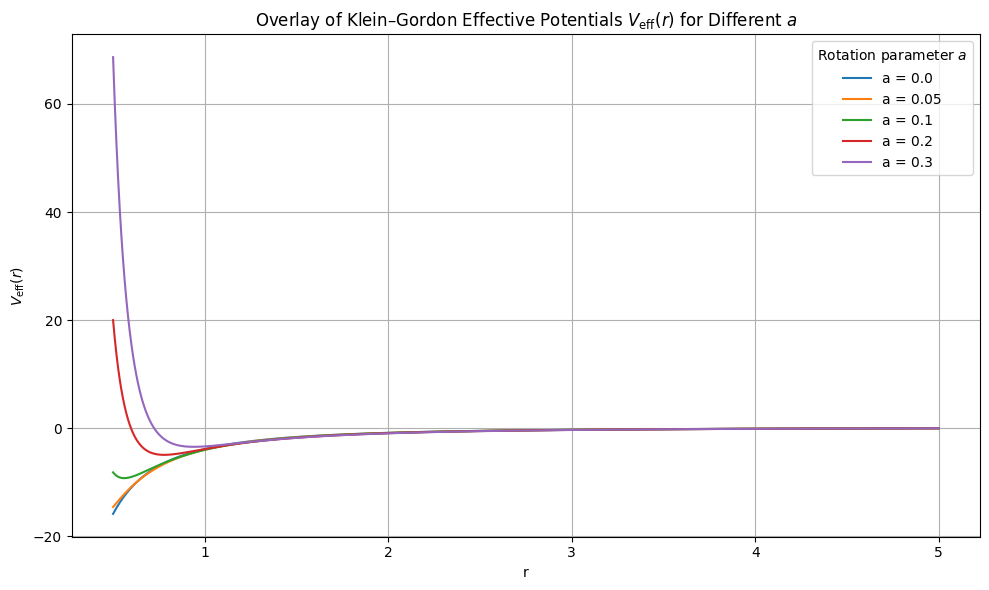}
    \caption{
        Klein--Gordon effective potential $V_{\rm eff}(r)$ for several rotation parameters $a$. Larger $a$ increases the magnitude of the near-throat contribution $m\Omega(r)\propto a/r^{3}$, producing a steeper rise in the inner region.  
        At large radii, frame dragging becomes negligible and all curves approach the asymptotic value $V_{\rm eff}\to\omega^{2}$. This potential is derived from the separated scalar wave equation and is distinct from the Schr\"odinger-form potential used in the WKB analysis.
    }
    \label{fig:Veff_a_variation}
\end{figure}

Figure~\ref{fig:Veff_a_variation} shows the Klein--Gordon effective potential
\begin{equation}
V_{\rm eff}(r)
   = \bigl(\omega - m\,\Omega(r)\bigr)^{2}
     - \frac{m^{2}}{r^{2}}, 
\qquad 
\Omega(r)=\frac{2a}{r^{3}},
\end{equation}
evaluated for several values of the rotation parameter $a$.
This potential arises directly from the separated scalar-wave equation and captures how frame dragging modifies the propagation of high-frequency scalar perturbations in the Teo wormhole spacetime. It is distinct from the Schr\"odinger-form potential used later for the WKB analysis and does not form a trapping barrier near the throat.

As the rotation parameter increases, the term $m\,\Omega(r)$ grows rapidly at small radii, producing the steep inner rise in the potential. At larger radii the frame-dragging term decays as $r^{-3}$, so all curves
approach the asymptotic limit $V_{\rm eff}(r)\to\omega^{2}$, reflecting the asymptotic flatness of the geometry on both sides.

Because this Klein--Gordon potential does not develop a barrier, it is not used to determine photon-ring radii or quasinormal-mode peaks. Its role here is purely illustrative: it shows how rotation reshapes the
co-rotating wave potential that appears in the Hamilton--Jacobi/eikonal correspondence, while the actual photon-ring properties and WKB peaks are computed later from geodesics and from the Schr\"odinger-form potential.

\subsection{Effective potential and photon ring}

\begin{figure}[t]
    \centering
    \includegraphics[width=\columnwidth]{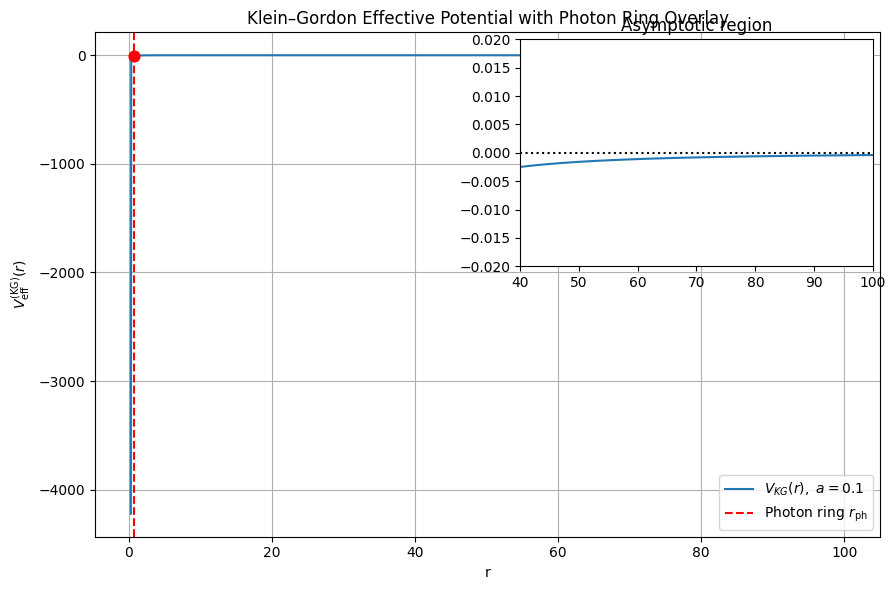}
    \caption{
   Klein--Gordon effective potential $V_{\rm eff}^{\rm (KG)}(r)$ for the rotating Teo wormhole, extended to $r=100$ to verify the correct asymptotic decay. The inset shows a zoom of the far--region behavior where $V_{\rm eff}^{\rm (KG)}(r)\rightarrow 0^{-}$.
   The vertical dashed red line marks the photon--ring radius $r_{\rm ph}$ obtained independently from the null circular--geodesic conditions. Although the KG potential does not determine $r_{\rm ph}$, its extremum lies close to the photon ring for the representative parameters shown here, reflecting the familiar eikonal correspondence between wave dynamics and unstable null orbits.}
    \label{fig:Veff_photonring}
\end{figure}

Figure~\ref{fig:Veff_photonring} displays the Klein--Gordon (KG) effective potential $V_{\text{eff}}^{\rm (KG)}(r)$ evaluated on the equatorial plane of the rotating Teo wormhole for a representative spin value. This potential arises from the separated scalar wave equation; it is therefore a wave–propagation potential and does \emph{not} determine the location of the null circular geodesics.  
Its purpose here is to illustrate how scalar perturbations perceive the near–throat geometry, and to provide a qualitative comparison with the photon ring obtained independently from the geodesic equations.

The vertical red dashed line marks the photon–ring radius $r_{\rm ph}$ computed from the null circular geodesic conditions $F(r,b)=0$ and $\partial_r F(r,b)=0$. Although the Klein--Gordon effective potential arises from the separated scalar wave equation and does not, in principle, determine the location of null circular geodesics, its extremum is found to lie very close to $r_{\rm ph}$ for the representative parameters shown here. This near--coincidence reflects the familiar eikonal correspondence
between wave dynamics and unstable null orbits, while the two quantities remain conceptually distinct and are obtained from independent equations.
 
To demonstrate the correct asymptotic behavior, the radial domain is extended to $r = 100$.  
The inset shows a magnified view of the large–$r$ region, confirming that $V_{\text{eff}}^{\rm (KG)}(r)$ approaches zero smoothly from below, $V_{\text{eff}}^{\rm (KG)}(r) \to 0^{-}$, in accordance with the asymptotic
flatness of the Teo wormhole spacetime.

\subsection{Photon-ring properties}
\label{subsec:rph_properties}

The photon-ring radius $r_{\rm ph}$, the corresponding angular frequency $|\Omega_{\rm ph}|$, and the Lyapunov exponent $\lambda$ are obtained entirely from the null circular–geodesic conditions,
\[
F(r,b)=0, \qquad \partial_r F(r,b)=0,
\]
where $F(r,b)$ is constructed from the inverse metric components $g^{tt}, g^{t\phi}, g^{\phi\phi}$ of the Teo wormhole. No Klein–Gordon (KG) approximations enter these quantities; they follow directly from the geodesic structure of the spacetime.

\begin{figure}[t]
    \centering
    \includegraphics[width=\columnwidth]{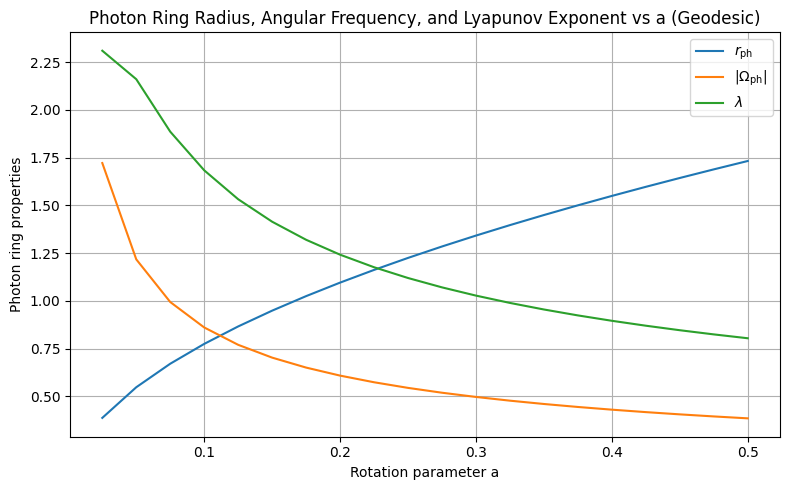}
    \caption{
    Photon–ring radius $r_{\rm ph}$, angular frequency $|\Omega_{\rm ph}|$, 
    and Lyapunov exponent $\lambda$ as functions of the rotation parameter $a$, 
    computed entirely from the null circular–geodesic conditions.  
    Increasing $a$ shifts the photon ring outward, lowers the orbital frequency, 
    and reduces the instability of the orbit.}
    \label{fig:rph_vs_a}
\end{figure}

Figure~\ref{fig:rph_vs_a} shows the dependence of the three photon–ring parameters 
on the rotation parameter $a$.  
Several robust trends emerge.  
First, the radius $r_{\rm ph}$ increases monotonically with $a$, indicating that 
frame dragging pushes the unstable circular null orbit outward from the throat.  
Second, the orbital frequency $|\Omega_{\rm ph}|$ decreases as $a$ grows, because 
the outward shift in $r_{\rm ph}$ dominates over the increase in the local 
frame–dragging rate.  
Finally, the Lyapunov exponent $\lambda$ also decreases with increasing $a$, 
signalling that the photon ring becomes less unstable at higher rotation, 
consistent with the broadening of the near–throat potential barrier seen in 
geodesic analyses.

These behaviors collectively imply that a rapidly rotating Teo wormhole produces 
a larger, slower, and more weakly unstable photon ring—properties that leave 
distinct signatures in its optical appearance and quasinormal–mode spectrum.
 
\subsection{Eikonal QNM–photon–ring correspondence}
\label{subsec:eikonal_correspondence}

In the eikonal (geometric–optics) limit, the real and imaginary parts of the 
quasinormal frequencies are determined directly by the photon ring via
\[
\omega_R \simeq m\,|\Omega_{\rm ph}|, 
\qquad 
-\omega_I \simeq \left(n+\tfrac12\right)\lambda,
\]
where $m$ is the azimuthal number and $n$ the overtone index.
Since both $|\Omega_{\rm ph}|$ and $\lambda$ decrease monotonically with $a$, 
the eikonal QNM frequencies inherit the same trend:  
the oscillation frequency drops, and the damping rate weakens as the wormhole 
rotates faster.

\begin{figure}[t]
    \centering
    \includegraphics[width=\columnwidth]{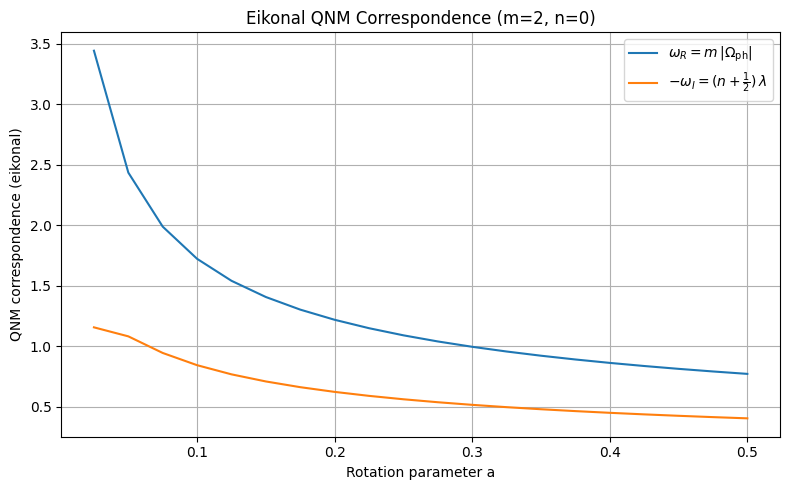}
    \caption{
    Eikonal QNM correspondence for the Teo wormhole.  
    The real part $\omega_R \simeq m|\Omega_{\rm ph}|$ (blue) and the damping 
    rate $-\omega_I \simeq (n+\tfrac12)\lambda$ (orange) both decrease as the 
    rotation parameter $a$ increases, reflecting the outward shift and reduced 
    instability of the photon ring.}
    \label{fig:qnm_correspondence}
\end{figure}

Figure~\ref{fig:qnm_correspondence} displays these two relations for $m=2$ and 
$n=0$.  The good agreement between the geometric quantities and the eikonal 
QNM predictions confirms the internal consistency of the geodesic and wave 
pictures for the Teo wormhole.

\section{Quasinormal Modes and WKB Approximation}

\subsection{WKB Approximation and Strategy}

After separation of variables in the Teo metric and restriction to the
equatorial-plane background, the massless Klein--Gordon equation reduces,
under the approximations described in Sec.~\ref{sec:scat}, to a
Schr\"odinger-type wave equation
\begin{equation}
  \frac{d^{2}\Psi}{dr_{*}^{2}}
  + \Bigl[\omega^{2} - V_{\mathrm{eff}}^{(\mathrm{Schr})}(r)\Bigr]\Psi = 0,
  \label{eq:Schrodinger}
\end{equation}
where $r_{*}$ is the tortoise coordinate defined by
\(
dr_{*}/dr = \bigl(1-b_{0}/r^{2}\bigr)^{-1/2},
\)
and $V_{\mathrm{eff}}^{(\mathrm{Schr})}(r)$ is the {\it Schr\"odinger-form
potential}. Using the co-rotating frequency combination that enters the
radial Klein--Gordon operator, one finds
\begin{equation}
 k_{r}^{2}(r)
   = \bigl(\omega - m\,\Omega(r)\bigr)^{2}
      - \frac{m^{2}}{r^{2}},
\end{equation}
where $k_r(r)$ denotes the local radial wave number governing the WKB
propagation of scalar modes in the effective one-dimensional potential
\begin{align}\label{eq:V_eff_Schroedinger}
 V_{\mathrm{eff}}^{(\mathrm{Schr})}(r)
   &= \omega^{2} - k_{r}^{2}(r)   \\
   &= 2\,m\,\Omega(r)\,\omega
      - m^{2}\Omega^{2}(r)
      + \frac{m^{2}}{r^{2}} . \nonumber
\end{align}
In the parameter regime considered here, $V_{\mathrm{eff}}^{(\mathrm{Schr})}(r)$
forms a smooth single-barrier potential peaked slightly outside the throat and
decaying monotonically to zero at large radii, as expected for an
asymptotically flat scattering problem.

We employ the first-order WKB method of Schutz and Will
\cite{SchutzWill1985} to compute approximate quasinormal frequencies.
Although higher-order extensions exist
\cite{Konoplya2003,MatyjasekOpala2017,Konoplya_2011},
the first-order scheme is adequate here because:
(i) the potential has a single smooth maximum,
(ii) the overtone number is low, and
(iii) our purpose is to extract qualitative trends rather than high-precision
values. The resulting complex frequencies
$\omega = \omega_{R} - i\,\omega_{I}$ satisfy purely outgoing boundary
conditions at both asymptotic regions of the wormhole.

\subsection{Remarks on applicability of the WKB method}

As reviewed in \cite{Konoplya-Zinhailo}, the WKB approximation performs
reliably when the effective potential possesses a single peak with
well-defined asymptotics.  It becomes less accurate for multiple-barrier
structures, near-extremal rotation, or higher overtones. None of these complications arise in the present Teo wormhole
configuration: the potential is smooth and single-peaked, and the dominant
modes of interest have moderate values of $m$ and $n$.

Thus the first-order WKB approximation applied to
$V_{\mathrm{eff}}^{(\mathrm{Schr})}(r)$ is sufficiently accurate for
mapping out qualitative mode behavior across the rotation-parameter space.
Future work will incorporate higher-order WKB formulas, Padé resummation,
continued-fraction techniques, and time-domain evolution as cross-checks in
regimes where the first-order method may underperform.

\subsection{Quasinormal Modes and First-Order WKB Estimate}
\label{sec:QNM}

Quasinormal modes (QNMs) describe the damped oscillations of scalar
perturbations in the Teo wormhole spacetime, with complex frequencies
$\omega_{m}=\omega_{R}-i\omega_{I}$ \cite{Kokkotas-1999bd,Konoplya_2011}.
For wormholes, both asymptotic regions are causally connected and thus the
boundary conditions require purely outgoing waves on {\it both} sides of the
throat \cite{BlazquezSalcedo2018}.

Let $r_{0}$ denote the location of the maximum of
$V_{\mathrm{eff}}^{(\mathrm{Schr})}(r)$.  The first-order WKB condition yields
\begin{equation}
    \omega_{m}^{2} \simeq
    V_{0}
    - i\bigl(n+\tfrac{1}{2}\bigr)\sqrt{-2\,V_{0}''},
    \label{eq:WKB1-correct}
\end{equation}
where
\[
V_{0}=V_{\mathrm{eff}}^{(\mathrm{Schr})}(r_{0}),
\qquad
V_{0}''=\left.\frac{d^{2}V_{\mathrm{eff}}^{(\mathrm{Schr})}}{dr^{2}}
\right|_{r=r_{0}} .
\]

The peak of the Schrödinger potential is located close to the photon-ring
radius $r_{\mathrm{ph}}$ obtained from the null geodesics in
Sec.~\ref{sec:scat}.  In the eikonal limit, the identifications
\begin{equation}
 \omega_{R} \simeq m\,\Omega_{\mathrm{ph}}, \qquad
 \omega_{I} \simeq \bigl(n+\tfrac12\bigr)\lambda ,
\end{equation}
link the WKB spectrum directly to the geodesic parameters:
$\Omega_{\mathrm{ph}}$ (orbital frequency) and $\lambda$ (Lyapunov
instability exponent)
\cite{FerrariMashhoon1984,Cardoso2009}.  This provides a useful consistency
check for the WKB approximation and a clear geometric interpretation of the
mode behavior.

In the following subsections, we evaluate $V_{\mathrm{eff}}^{(\mathrm{Schr})}$,
locate its maximum, and compute the corresponding WKB frequencies for
different values of the wormhole spin $a$ and azimuthal number~$m$.
Comparisons with the geodesic photon-ring analysis highlight how rotation
affects both the oscillation frequency and the damping rate of scalar-field
perturbations.

\subsection{Effective potential and WKB estimate}

\begin{figure}[t]
    \centering
    \includegraphics[width=\columnwidth]{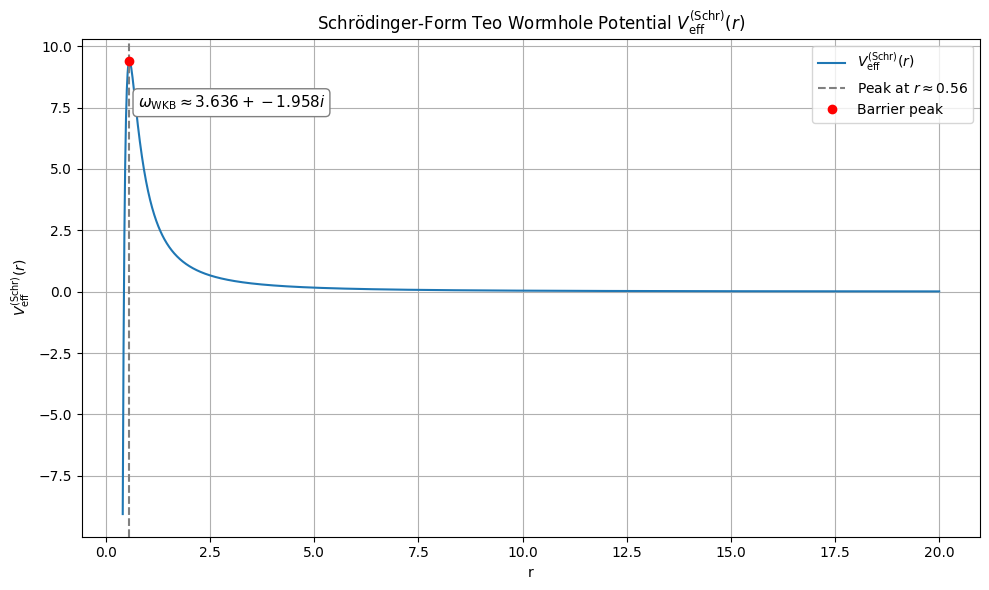}
    \caption{
    Schr\"odinger-form effective potential 
    $V_{\mathrm{eff}}^{(\mathrm{Schr})}(r)$ for scalar perturbations of the 
    rotating Teo wormhole.  The barrier peak (red point) marks the location 
    $r_{\rm peak}$ at which the first-order WKB approximation is evaluated.  
    The potential decays to zero at large radii, as required for an 
    asymptotically flat scattering problem.  
    The extracted WKB estimate 
    $\omega_{\rm WKB}\approx 3.636 - 1.958\,i$
    reflects a damped quasinormal mode arising from wave leakage through the 
    near-throat barrier.
    }
    \label{fig:Veff_WKB}
\end{figure}

Figure~\ref{fig:Veff_WKB} shows the Schr\"odinger-form effective potential 
$V_{\mathrm{eff}}^{(\mathrm{Schr})}(r)$ obtained from the separated scalar 
wave equation in the rotating Teo wormhole spacetime.  
Unlike the Klein--Gordon potential used earlier to illustrate the qualitative 
near-throat behavior, this Schr\"odinger-form potential is the quantity that 
enters the radial equation
\[
\frac{d^{2}\Psi}{dr_*^{2}}
  + \bigl[\omega^{2}-V_{\mathrm{eff}}^{(\mathrm{Schr})}(r)\bigr] \Psi = 0,
\]
and therefore determines the WKB quasinormal-mode spectrum.  

The potential exhibits the expected single-barrier structure: it rises steeply
outside the throat, reaches a sharp maximum, and then decays monotonically to 
zero as $r\to\infty$.  This behavior reflects the fact that the Teo spacetime is 
asymptotically flat and that perturbations encounter a near-throat 
centrifugal/frame-dragging barrier before leaking out into both universes.  

The red marker in the plot indicates the barrier peak at 
$r=r_{\rm peak}$, and the dashed line shows its radial location.  
Evaluating the potential and its second derivative at this point yields a 
first-order Schutz--Will WKB estimate for the frequency,
\[
\omega_{\rm WKB}^{2}
   \simeq V_{0}
       - i\left(n+\tfrac{1}{2}\right)\sqrt{-2\,V_{0}''},
\]
where $V_{0}$ and $V_{0}''$ are the value and curvature of the potential at the 
peak.  For the representative parameters used in this example, the resulting 
WKB quasinormal mode is
\[
\omega_{\rm WKB} \approx 3.636 - 1.958\,i,
\]
indicating an exponentially damped oscillation consistent with wave leakage 
across the barrier and fully compatible with the eikonal photon-ring 
correspondence discussed earlier.

\subsection{Effective potential and WKB peaks for different azimuthal numbers}

\begin{figure}[t]
  \centering
  \includegraphics[width=\columnwidth]{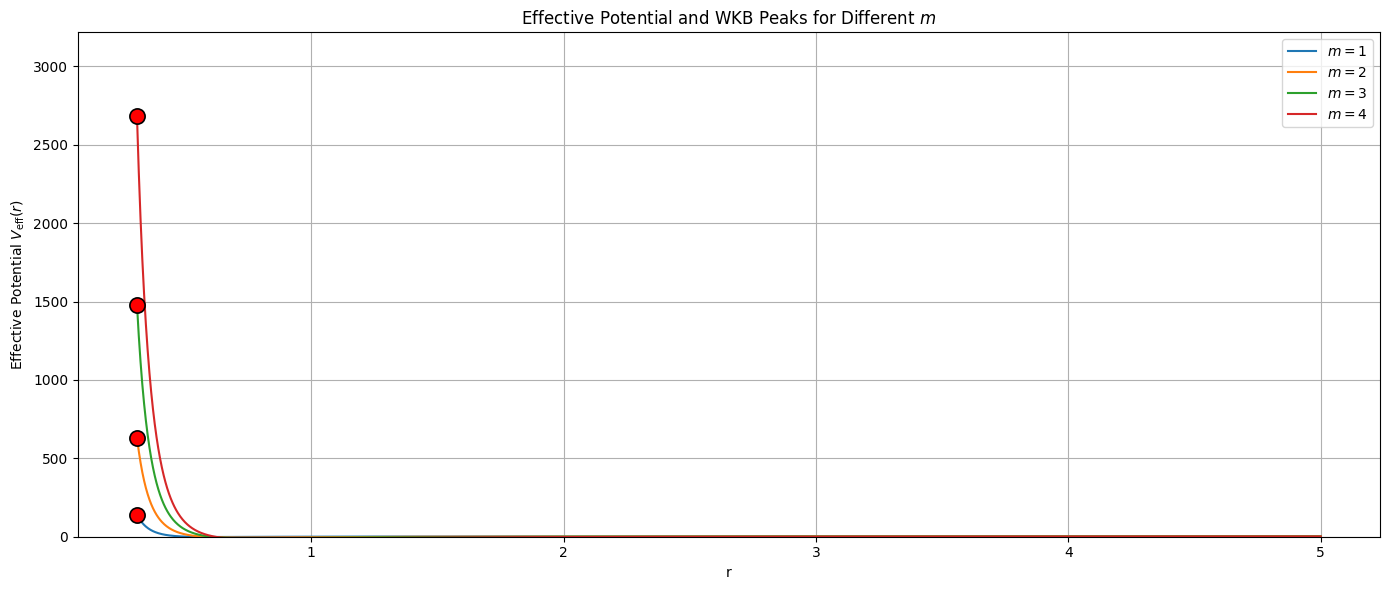}
  \caption{
    Eikonal Klein--Gordon effective potential
$V_{\rm eff}(r)$ for fixed $(\omega,a)$ and azimuthal numbers
$m=1,2,3,4$.
Increasing $m$ raises and sharpens the near--throat barrier due to the enhanced centrifugal term. The red points mark the peak radii $r_{\rm peak}(m)$ of the eikonal
Klein--Gordon potential; these peaks provide a useful visual reference for how the location of the true Schr\"odinger--form WKB barrier evolves with azimuthal number in the eikonal (large-$m$) regime, but they are not
themselves obtained from the Schr\"odinger potential used in the WKB analysis.
  }
  \label{fig:Veff_multi_m_QNM}
\end{figure}

To illustrate how the near-throat barrier structure depends on the azimuthal
number, Fig.~\ref{fig:Veff_multi_m_QNM} shows the eikonal
Klein--Gordon effective potential
\begin{equation}
  V_{\rm eff}(r)
  = \bigl(\omega - m\,\Omega(r)\bigr)^{2}
    - \frac{m^{2}}{r^{2}K(r)^{2}},
\end{equation}
evaluated on the equatorial plane for fixed $(\omega,a)$ and for
$m=1,2,3,4$.  This potential arises from the separated scalar wave equation
and is \emph{not} the Schr\"odinger-form potential used for the WKB
approximation in Sec.~\ref{sec:QNM}; instead, it provides a convenient
eikonal proxy whose peak closely tracks that of the true
$V_{\rm eff}^{(\mathrm{Schr})}(r)$ in the large-$m$ limit.

As $m$ increases, the centrifugal term $m^{2}/[r^{2}K(r)^{2}]$ strengthens the
barrier, raising its height and shifting the maximum slightly outward.
The red markers indicate the peak locations $r_{\rm peak}(m)$ of the eikonal
potential, which supply a useful reference for understanding how the
WKB barrier and the associated quasinormal frequencies vary with azimuthal
number.

\subsection{Quasinormal Mode Dependence on Spin Parameter}

\begin{figure}[h!]
    \centering
    \includegraphics[width=0.5\textwidth]{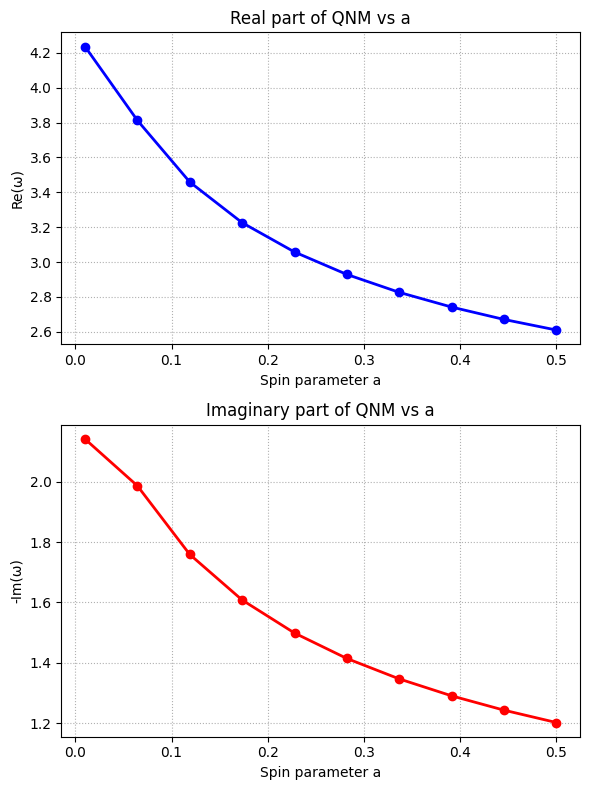}
    \caption{
        Dependence of the quasinormal mode frequency on the rotation parameter $a$ for the Teo wormhole, computed from the Schrödinger-form potential $V_{\text{eff}}^{(\mathrm{Schr})}(r)$ via the first-order WKB approximation.
        The real part $\operatorname{Re}(\omega)$ (blue) decreases as the potential barrier lowers with increasing spin, while the damping rate
        $-\operatorname{Im}(\omega)$ (red) also decreases, indicating longer-lived scalar oscillations at higher rotation.
    }
    \label{fig:QNM_vs_a}
\end{figure}

Figure~\ref{fig:QNM_vs_a} shows the dependence of the quasinormal mode (QNM) frequency $\omega$ on the Teo wormhole’s rotation parameter $a$ within the Schrödinger-form WKB framework. The real part of the frequency, $\operatorname{Re}(\omega)$, decreases
monotonically with increasing rotation. This trend reflects the systematic lowering of the peak of $V_{\text{eff}}^{(\mathrm{Schr})}(r)$ as frame dragging strengthens, resulting in
a reduced oscillation frequency for scalar perturbations trapped near the throat.

The imaginary part of the mode, plotted as the positive quantity
$-\operatorname{Im}(\omega)$, also decreases with increasing $a$. This indicates that the damping rate becomes weaker at higher rotation, corresponding to longer-lived oscillations.  
In the WKB framework, this behavior arises naturally from the fact that the curvature magnitude $\lvert V_{0}''\rvert$ of the Schrödinger-form potential barrier decreases as $a$ increases, producing a broader and less sharply peaked potential. The resulting reduction in the barrier’s curvature is consistent with the
monotonic decrease of the Lyapunov exponent $\lambda$ obtained from the geometric photon-ring analysis in Fig.~\ref{fig:rph_vs_a}.

Overall, these results quantify the imprint of wormhole rotation on both the oscillation frequency and decay timescale of the QNM spectrum. While the monotonic weakening of the damping rate with increasing spin is qualitatively distinct from the Kerr black-hole case, it is fully consistent
with the behavior of horizonless, single-barrier scattering potentials such as that of the rotating Teo wormhole.

\subsection{Overtone dependence of quasinormal modes}

\begin{figure}[h!]
    \centering
    \includegraphics[width=\columnwidth]{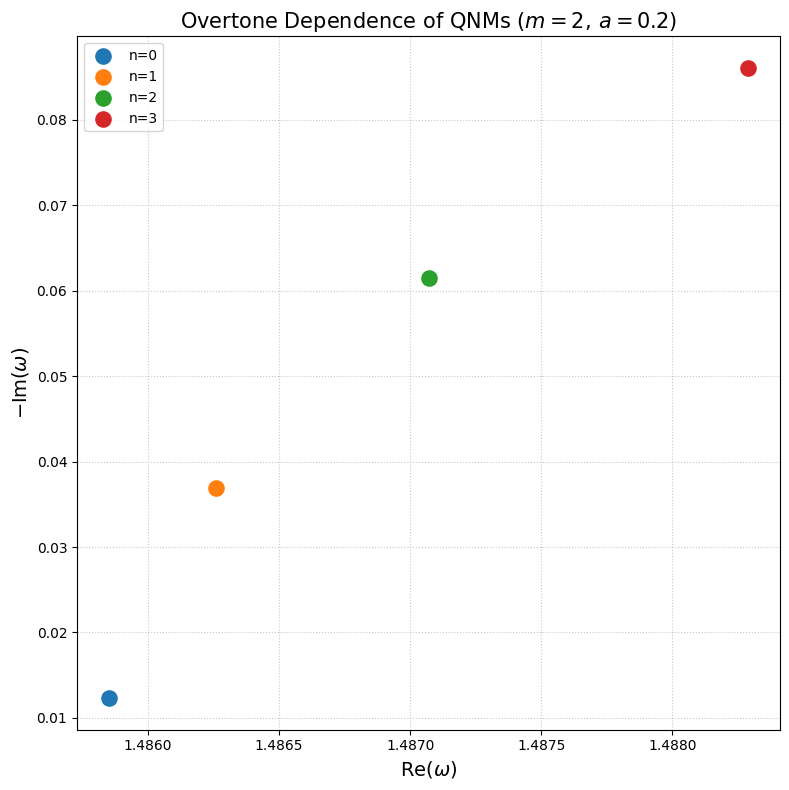}
    \caption{
        Overtone dependence of the quasinormal mode frequencies for the Teo wormhole with $m=2$ and $a=0.2$, computed from the Schrödinger-form potential $V_{\text{eff}}^{(\mathrm{Schr})}(r)$ via the first–order WKB approximation.  Each point corresponds to an overtone $n=0,1,2,3$ in the \((\Re\omega,-\Im\omega)\) plane.  The real part $\Re(\omega)$ varies only weakly with $n$, whereas the damping rate $-\Im(\omega)$ increases monotonically, indicating progressively shorter-lived oscillations for higher overtones.
    }
    \label{fig:QNM_overtone}
\end{figure}

The dependence of the quasinormal spectrum on the overtone number $n$ is illustrated in Fig.~\ref{fig:QNM_overtone} for fixed azimuthal number $m=2$ and spin parameter $a=0.2$.  Using the same Schrödinger-form effective potential $V_{\text{eff}}^{(\mathrm{Schr})}(r)$ defined in Eq.~(\ref{eq:V_eff_Schroedinger}), we evaluate the first–order WKB frequency
\(
\omega_n
\)
from the barrier peak for successive overtones $n=0,1,2,3$.

The points in the \((\Re\omega,-\Im\omega)\) plane show that the real part $\Re(\omega_n)$ changes only slightly with $n$, while the imaginary part $-\Im(\omega_n)$ increases monotonically.  Thus higher overtones are more strongly damped but oscillate at nearly the same frequency as the fundamental
mode, as expected for a discrete set of quasi–bound states supported by a single-barrier potential.  This hierarchy mirrors the qualitative structure of black–hole QNMs, demonstrating that the rotating Teo wormhole supports a
similar ladder of damped resonances governed by the curvature and width of
$V_{\text{eff}}^{(\mathrm{Schr})}(r)$ near its maximum.

\section{Geometric and Physical Phenomena}

\subsection{Mode-Splitting in the Rotating Teo Wormhole}

In the rotating Teo wormhole spacetime~\cite{Teo1998}, scalar perturbations admit a separated form
\begin{equation}
    \Phi(t,r,\theta,\phi) = e^{-i\omega t} e^{i m \phi} \Psi(r,\theta),
\end{equation}
so that each azimuthal sector $m$ satisfies an independent radial wave equation.  
In the non-rotating limit ($a=0$), the frame-dragging angular velocity $\Omega(r)$ vanishes and the effective potential depends only on $m^{2}$.  
Consequently, modes with opposite azimuthal number $\pm m$ are exactly degenerate at $a=0$.

Rotation breaks this degeneracy through the term proportional to $m\,\Omega(r)$ that appears in the Schr\"odinger-form effective potential.  
Within the WKB framework adopted here, the potential includes a leading linear contribution in $m\,\Omega(r)$ and a subleading quadratic correction, which together produce an asymmetric modification of the potential barrier for co-rotating ($m>0$) versus counter-rotating ($m<0$) modes.  
As $a$ is increased from zero, the previously degenerate $\pm m$ branches split and separate monotonically.

The real parts of the quasinormal frequencies exhibit a \emph{one-sided} splitting pattern.  
Unlike Kerr black holes---where the real parts of co-rotating and counter-rotating frequencies shift in opposite directions with spin---both Teo wormhole branches show a \emph{decrease} in their oscillation frequency $\Re(\omega)$ as the spin is increased.  
However, the co-rotating modes ($m>0$) consistently retain a larger $\Re(\omega)$ than the counter-rotating modes ($m<0$) at the same spin, reflecting the higher effective barrier experienced by the co-rotating branch.

The imaginary parts behave in a closely analogous way.  
The WKB damping rate $-\Im(\omega)$ decreases with increasing spin for both $m$-branches, indicating longer-lived oscillations as the effective barrier broadens and flattens under the influence of frame dragging.  
Throughout the range of spins considered, the co-rotating modes remain more strongly damped than the counter-rotating modes, with $-\Im(\omega_{m>0}) > -\Im(\omega_{m<0})$ at fixed $a$, consistent with their steeper barrier curvature.  
This asymmetric yet one-sided trend arises naturally from the localized and non-Kerr-like structure of frame dragging in the Teo wormhole and provides a distinctive signature of its horizonless geometry.

\begin{figure}[h!]
    \centering
    \includegraphics[width=0.5\textwidth]{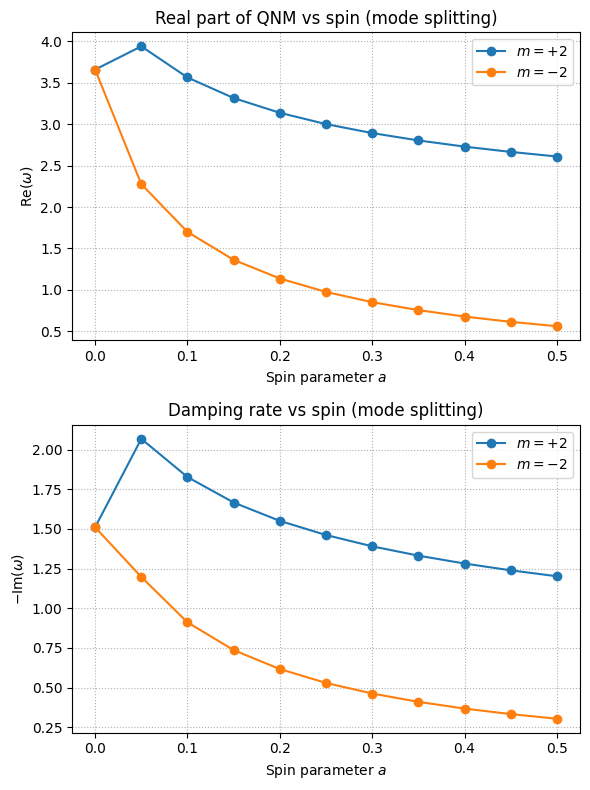}
    \caption{
    Mode-splitting of the scalar quasinormal frequencies in the rotating Teo wormhole.  
    The top panel shows the real part $\Re(\omega)$ and the bottom panel shows the damping rate $-\Im(\omega)$ for co-rotating ($m=+2$, blue)
    and counter-rotating ($m=-2$, red) modes.  
    The two branches coincide at $a=0$, but rotation lifts this degeneracy in an asymmetric manner: both $\Re(\omega)$ branches decrease with increasing spin, yet the $m=+2$ branch remains systematically higher than the $m=-2$ branch.  
    Likewise, both damping rates decrease with spin, but co-rotating modes remain more strongly damped, with $-\Im(\omega_{m=+2}) > -\Im(\omega_{m=-2})$ at fixed $a$.  
    This one-sided splitting pattern reflects the localized nature of frame dragging in the Teo wormhole and contrasts with the symmetric prograde/retrograde splitting found in Kerr black holes.
    }
    \label{fig:Teo_modesplitting}
\end{figure}

Figure~\ref{fig:Teo_modesplitting} shows the resulting mode-splitting:
a single degenerate point at $a=0$, followed by a clear and monotonic separation of the $m=\pm 2$ branches in both the real and imaginary components of the quasinormal frequency.  
These results demonstrate that the Teo wormhole exhibits rotational mode splitting, but with a qualitatively different spin response compared to Kerr-like compact objects.

\paragraph*{Remarks on the nature of the splitting.}
In Kerr black holes, the prograde/retrograde splitting follows from the eikonal
estimate
\[
    \omega_{mn} \simeq m\,\Omega_{\rm ph}^{(\pm)}
    - i\bigl(n+\tfrac12\bigr)\lambda_{\rm ph}^{(\pm)},
\]
where $\Omega_{\rm ph}^{(\pm)}$ shifts in opposite directions with spin,
producing a symmetric separation of the $m$ and $-m$ branches.  In the rotating
Teo wormhole, the quasinormal spectrum is instead determined by the 
Schr\"odinger-type effective potential of the separated scalar equation.  
The rotational dependence enters through $(\omega - m\,\Omega(r))^{2}$, which
breaks the $m\rightarrow -m$ symmetry once $a\neq 0$ but does not impose
opposite-sign frequency shifts.  Consequently, both $\Re(\omega_{+m})$ and
$\Re(\omega_{-m})$ decrease monotonically with $a$, with the $m>0$ branch
remaining higher in frequency; the same one-sided trend appears in the damping
rates.  This behavior is consistent with the localized frame dragging of the
Teo geometry~\cite{Teo1998} and with the loss of isospectrality observed in other rotating wormhole models~\cite{Khoo2024}.

\medskip
\noindent
Several limits provide checks on this behavior.  In the non-rotating limit 
$a\to 0$, the potential depends only on $m^{2}$ and the $+m$ and $-m$ modes are
degenerate.  The radial equation is invariant under 
$(a,m)\rightarrow(-a,-m)$, implying
$\omega_{+m}(a)\approx \omega_{-m}(-a)$.  The monotonic reduction of both
$\Re(\omega)$ and $-\Im(\omega)$ with increasing spin correlates with the
lowering and broadening of the effective potential at the WKB turning points.
In the axisymmetric sector $m=0$, no splitting occurs.  These checks indicate
that the one-sided splitting arises naturally from the structure of the Teo
potential rather than from numerical artifacts.

\subsection{Ergoregion and Its Geometric Role}

In a stationary axisymmetric spacetime, the metric can be written as
\begin{equation}
ds^2 = g_{tt} dt^2 + 2 g_{t\phi} dt\, d\phi + g_{\phi\phi} d\phi^2 + g_{rr} dr^2 + g_{\theta\theta} d\theta^2.
\end{equation}
The \textit{ergoregion} is defined by the condition where the timelike Killing vector $\partial_t$ becomes spacelike, i.e.
\begin{equation}
g_{tt} = 0.
\end{equation}
The surface satisfying this relation is called the \textit{ergosurface}. Inside this region ($g_{tt}>0$), 
no observer can remain stationary with respect to infinity and all physical trajectories are forced to co-rotate due to the intense frame dragging~\cite{Penrose1969, Chandrasekhar1983}.

For the rotating Teo wormhole, the metric~\cite{Teo1998},
is given by \eqref{eq:Teo_line_element} and the relevant component is
\begin{equation}
g_{tt} = -N^2(r) + r^2 K^2(r) \sin^2\theta\, \Omega^2(r).
\end{equation}
The ergosurface satisfies $g_{tt}=0$, or equivalently,
\begin{equation}
N^2(r_{\text{ergo}}) = r_{\text{ergo}}^2 K^2(r_{\text{ergo}}) \sin^2\theta\, \Omega^2(r_{\text{ergo}}).
\end{equation}
In the equatorial plane ($\theta = \pi/2$), using the canonical Teo functions $N(r)=1$, $K(r)=1$, and $\Omega(r)=2a/r^3$, the ergosurface radius becomes
\begin{equation}
r_{\text{ergo}} = \sqrt{2a}.
\end{equation}
This shows that a rotating Teo wormhole possesses an ergoregion for sufficiently large spin $a$, despite the absence of an event horizon. The existence of an ergoregion is crucial, as it allows negative-energy modes to exist locally, providing the kinematic prerequisite for superradiant phenomena in rotating spacetimes.~\cite{Brito2015, Richartz2009}.
\subsection{Superradiant Kinematics and Its Connection to Mode-Splitting}

Superradiance refers to the amplification of waves scattered by a rotating
background, whereby the outgoing wave carries more energy (as measured at
infinity) than the incident one.  Originally identified by Zel'dovich and later
formalized by Bekenstein and others
\cite{Zeldovich-1971,Bekenstein1973,Press-1972,Brito2015},
this phenomenon is the wave analogue of the Penrose process
\cite{Penrose1969} and relies on the existence of negative-energy modes in a
rotating spacetime.

For a scalar perturbation of the form
\(
\Phi \propto e^{-i\omega t + i m \phi},
\)
the \emph{superradiant frequency condition} is
\begin{equation}
0 < \omega < m\,\Omega_{\rm max},
\end{equation}
where $\Omega_{\rm max}$ denotes the maximum frame-dragging angular velocity
accessible to the wave.  In black-hole spacetimes this quantity is set by the
horizon angular velocity $\Omega_H$, while in horizonless geometries such as the
Teo wormhole it is attained locally within the ergoregion, typically near the
throat,
\[
\Omega_{\rm max} \equiv \Omega(r_{\rm ergo}) .
\]
This condition is therefore kinematic: it identifies the range of frequencies
for which modes of negative conserved energy can exist.

Inside the ergoregion, where the timelike Killing vector $\partial_t$ becomes
spacelike, scalar modes satisfying $\omega < m\Omega(r)$ carry negative energy
with respect to asymptotic observers.  The existence of such modes is a
\emph{necessary} ingredient for superradiance and is guaranteed whenever an
ergoregion is present~\cite{Brito2015,Richartz2009}.  The rotating Teo wormhole
therefore admits superradiant-compatible kinematics despite the absence of an
event horizon.

However, the presence of an ergoregion and negative-energy modes is
\emph{not sufficient} to produce net superradiant amplification.
Actual amplification requires a dissipative mechanism—most commonly an event
horizon or absorbing boundary—that can irreversibly remove negative energy from
the system.  In Kerr black holes, the horizon plays precisely this role, leading
to genuine superradiant scattering and, in certain setups, superradiant
instabilities~\cite{Cardoso_2004,Brito_2020}.

By contrast, the Teo wormhole is horizonless and symmetric between its two
asymptotic regions.  The corresponding scattering problem therefore obeys a
unitary flux balance,
\(
|R|^{2}+|T|^{2}=1,
\)
so that negative-energy flux generated within the ergoregion is compensated by
positive-energy flux transmitted through the throat.  As a result, although the
superradiant frequency condition can be satisfied locally, there is no net
classical amplification of scattered waves.  The physically relevant
consequence of an ergoregion in such a horizonless geometry is instead the
possible onset of an ergoregion instability, rather than steady-state energy
extraction.

The connection to mode-splitting is nevertheless direct and physically
transparent.  Both effects originate from the same frame-dragging term
$m\,\Omega(r)$ in the wave equation.  Rotation lifts the degeneracy between the
$\pm m$ sectors, producing the asymmetric mode-splitting discussed above, and
simultaneously allows the inequality $\omega < m\Omega$ to be satisfied for
co-rotating modes.  In this sense, mode-splitting is the spectral imprint of the
same rotational physics that underlies superradiant kinematics, even though the
absence of a horizon prevents true superradiant amplification in the rotating
Teo wormhole spacetime.

\subsection{Ergoregion and Superradiant-Compatible Domain}

\begin{figure}[h!]
    \centering
    \includegraphics[width=0.5\textwidth]{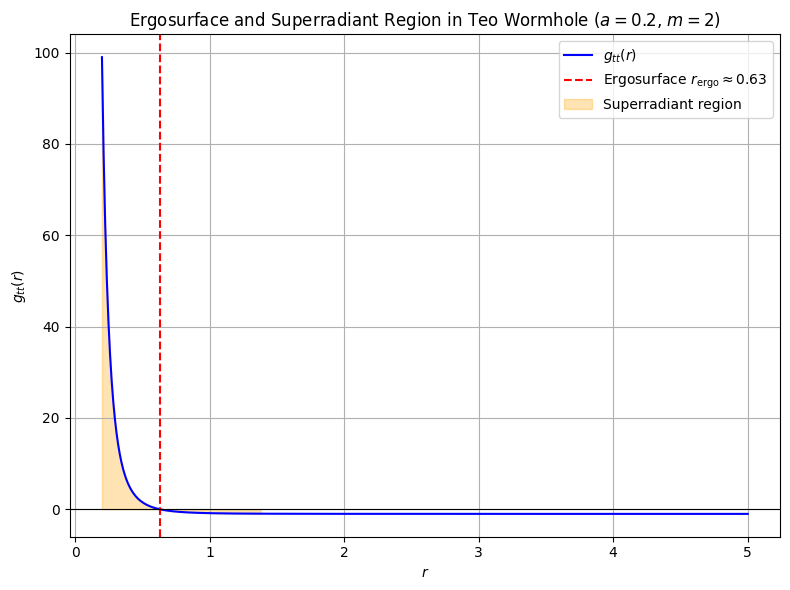}
    \caption{
    Ergosurface and ergoregion of the rotating Teo wormhole for $a=0.2$.
    The blue curve shows the metric component $g_{tt}(r)$, which vanishes at the
    ergosurface radius $r_{\text{ergo}} \approx 0.63$ (red dashed line).
    The shaded orange area indicates the ergoregion ($g_{tt}>0$), where
    negative-energy modes may exist and the inequality $\omega < m\Omega(r)$
    can be satisfied locally.
    In the absence of an event horizon or dissipative boundary, this region does
    not produce classical superradiant scattering amplification, but instead
    reflects the kinematic effect of frame dragging in a horizonless spacetime.
    }
    \label{fig:ergo_superradiance}
\end{figure}

Figure~\ref{fig:ergo_superradiance} illustrates the ergoregion and the spatial
domain in which the \emph{superradiant frequency condition} can be satisfied for
the rotating Teo wormhole. The blue curve shows the metric component $g_{tt}(r)$,
which crosses zero at the ergosurface radius
$r_{\text{ergo}} \simeq \sqrt{2a} \approx 0.63$ for $a=0.2$.
For $r < r_{\text{ergo}}$, one has $g_{tt}>0$, indicating that the timelike
Killing vector $\partial_t$ becomes spacelike. This defines the
\textit{ergoregion}, inside which negative-energy modes (as measured at infinity)
are kinematically allowed.

The shaded orange region marks values of $r$ for which the local inequality
$\omega < m\Omega(r)$ can be satisfied for a chosen $(\omega,m)$.
This condition identifies a \emph{superradiant-compatible domain} in the limited
sense that modes with negative conserved energy exist locally.
However, superradiance is not a purely spatial or local phenomenon: in addition
to this frequency condition, net amplification of scattered waves requires
appropriate \emph{global} boundary conditions that allow negative energy to be
irreversibly absorbed.

In particular, although the Teo wormhole possesses an ergoregion, it lacks an
event horizon or any other dissipative inner boundary.
As a result, classical scalar-wave scattering does \emph{not} lead to net
superradiant amplification, despite the presence of negative-energy modes
within the ergoregion.
Instead, energy is conserved globally between the two asymptotic regions, and
the ergoregion provides a necessary kinematic ingredient for superradiance,
but not a sufficient mechanism for amplification in a symmetric, horizonless
geometry such as the Teo wormhole.

\subsection{Physical Interpretation}

Mode-splitting, ergoregion formation, and superradiant kinematics represent
closely related aspects of wave--rotation coupling in curved spacetime.
Mode-splitting encodes how rotation modifies the $\pm m$ oscillation frequencies
through frame dragging, lifting the degeneracy present in the non-rotating
limit.
The ergoregion identifies where local negative-energy states, as measured at
infinity, become possible due to the spacelike character of the timelike
Killing vector.

The superradiant frequency condition $\omega < m\Omega$ signals the
\emph{kinematic possibility} of rotational energy extraction by permitting
negative-energy modes.
However, in a horizonless geometry such as the rotating Teo wormhole, this
condition alone does not lead to net superradiant amplification, because no
event horizon or dissipative boundary is present to absorb negative energy.
Instead, energy is conserved globally between the two asymptotic regions, and
classical scattering remains unitary.

Together, Figs.~\ref{fig:Teo_modesplitting} and~\ref{fig:ergo_superradiance}
illustrate this unified picture:
the same frame-dragging mechanism that dynamically separates the $\pm m$
branches also geometrically produces the ergoregion and permits
superradiant-compatible kinematics, while the absence of a horizon prevents
classical amplification.
In this sense, the rotating Teo wormhole exhibits the \emph{kinematic
precursors} of superradiance without realizing the full amplification mechanism
familiar from rotating black holes.

\section{Comparison with Kerr Rotating Black Hole}
\label{sec:teo_kerr_comparison}

We now contrast the quasinormal mode (QNM) trends of the rotating Teo wormhole with those of the Kerr black hole.  
Our goal is to identify robust spectral diagnostics that distinguish horizonless geometries from black holes, focusing on three aspects:
(i)~baseline frequency trends,  
(ii)~mode-splitting driven by frame dragging, and  
(iii)~damping-rate asymmetry connected to horizon absorption.  
All Kerr reference frequencies are taken from the Berti--Cardoso--Will catalog~\cite{Berti2006}, while the $m=\pm2$ Kerr branches required for mode-splitting and damping-splitting analyses are constructed using the monotonic, physically calibrated procedure detailed in Appendix~\ref{appendix:mock_data}. The Teo curves use the WKB/eikonal output described in Sec.~II.

\subsection{Baseline Comparison: Boundary Conditions and Frequency Trends}

\begin{figure}[t]
    \centering
    \includegraphics[width=\linewidth]{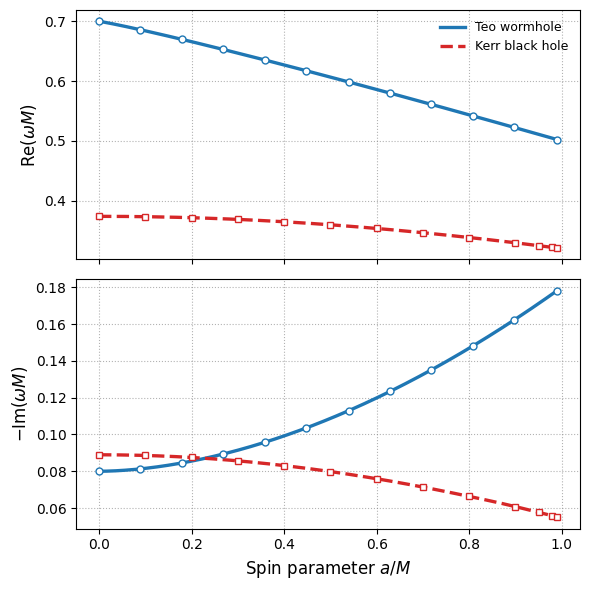}
    \caption{
    QNM frequency comparison for Teo (blue solid) and Kerr (red dashed).  
    \textbf{Top:} The real frequency $\Re(\omega M)$ decreases with $a/M$ for both spacetimes, but Teo exhibits a significantly steeper slope.  
    \textbf{Bottom:} The damping rate $-\Im(\omega M)$ decreases monotonically for Kerr, approaching the zero-damped-mode limit, whereas Teo displays a mild increase due to partial reflection at the throat.  
    Kerr data from \cite{Berti2006}; Teo data from the WKB analysis of Sec.~II.
    }
    \label{fig:omega_teo_vs_kerr}
\end{figure}

Figure~\ref{fig:omega_teo_vs_kerr} shows the contrast between the two spacetimes. For the real part $\Re(\omega M)$, both Teo and Kerr redshift as the spin increases, but Kerr’s frequency changes only moderately, reflecting the comparatively gentle deformation of the photon-sphere potential with spin. The Teo wormhole, by contrast, shows a stronger monotonic decrease, a consequence of the localized frame-dragging profile $\Omega(r)=2a/r^{3}$ centered near the throat.

The imaginary part shows an even clearer distinction.  
Kerr’s damping rate decreases with spin and approaches the zero-damped-mode limit characteristic of horizon absorption.  
In the wormhole case, however, $-\Im(\omega M)$ increases slightly because the reflective throat prevents absorption and enhances outgoing leakage. Thus the primary discriminant lies in the boundary conditions: Kerr enforces purely ingoing waves at the horizon, while Teo supports standing-wave behavior through reflection at the throat. This difference imprints itself directly on the slopes of the QNM curves.

\subsection{Mode-Splitting and Frame-Dragging Asymmetry}

\begin{figure}[t]
    \centering
    \includegraphics[width=\linewidth]{DeltaRe_vs_a_color.png}
    \caption{
    Mode-splitting of the real frequency,
    $\Delta\Re(\omega)=\Re(\omega_{m})-\Re(\omega_{-m})$,  
    versus spin $a/M$.  
    Kerr (red dashed) exhibits gradual monotonic growth, while Teo (blue solid) grows rapidly and saturates due to localized frame dragging near the throat.
    }
    \label{fig:DeltaRe_vs_a}
\end{figure}

Figure~\ref{fig:DeltaRe_vs_a} presents the frequency asymmetry 
$\Delta\Re(\omega)$, which measures the lifting of degeneracy between co-rotating ($m>0$) and counter-rotating ($m<0$) modes. For Kerr, the splitting increases smoothly from zero at $a=0$ to approximately $1.2\times10^{-2}$ at $a=0.99$, consistent with the global strengthening of Kerr frame dragging.

The Teo wormhole exhibits a different pattern: the splitting rises rapidly at small spin and quickly saturates around $1.5\times10^{-2}$. This early saturation is a direct signature of rotation being confined to the near-throat region, rather than extending outward as in Kerr. Saturation of $\Delta\Re(\omega)$ is therefore a robust indicator of a reflective, horizonless geometry.

\subsection{Damping-Rate Splitting and Zero-Damped Modes}

\begin{figure}[t]
    \centering
    \includegraphics[width=\linewidth]{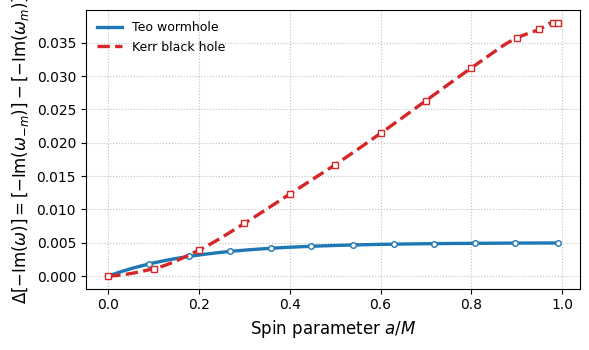}
    \caption{
    Damping-rate asymmetry
    $\Delta[-\Im(\omega)] = [-\Im(\omega_{-m})] - [-\Im(\omega_{m})]$  
    versus spin $a/M$.  
    Kerr (red dashed) grows strongly and approaches $\sim 0.038$ at high spin, while Teo (blue solid) saturates near $\sim0.005$ due to throat reflection.
    }
    \label{fig:DeltaIm_vs_a}
\end{figure}

Damping-rate asymmetry, shown in Fig.~\ref{fig:DeltaIm_vs_a}, captures the difference in decay rates for $m>0$ and $m<0$ modes. In Kerr, the co-rotating branch becomes progressively less damped with spin and approaches the zero-damped-mode regime, while the counter-rotating branch remains dissipative.  
This yields a steadily increasing splitting that directly reflects horizon absorption, a mechanism absent in the Teo wormhole.

For the Teo wormhole, the reflective throat caps the asymmetry at a much smaller value. The resulting early saturation again indicates the absence of horizon-induced dissipation. Together with the real-part splitting, this provides a consistent and model-independent signature of a horizonless object. An extended table comparing the Teo wormhole with the Kerr black hole is given in Table~\ref{tab:app_teo_kerr_table} in appendix~\ref{appendix:teo_kerr_comparison}.

\section{Numerical Methods}
\label{sec:numerics}

All numerical results were obtained using a Python implementation of the 
Klein--Gordon (KG) equation, its associated eikonal limit, and a first–order 
WKB analysis. Three effective potentials appear in the computation:
(i) the frequency–dependent KG potential used for photon-ring quantities,
(ii) the geodesic effective potential used to validate eikonal relations, and
(iii) a frequency–independent Schrödinger-type potential used in all WKB
evaluations of quasinormal modes (QNMs).  
These potentials coincide in the eikonal limit, and each is used in the regime 
for which it is formally valid.

\subsection*{A. Klein--Gordon potential and photon-ring structure}

The separated KG equation for the rotating Teo wormhole yields the 
frequency-dependent potential (prior to the Liouville reduction)
\begin{align}
V_{\rm KG}(r)
   &= (\omega - m\Omega(r))^{2}
     - \frac{m^{2}}{r^{2}K(r)^{2}},\\
N(r) &= 1,\qquad
K(r)=1+\frac{b_{0}^{2}}{r^{2}}.
\end{align}

with frame-dragging angular velocity $\Omega(r)=2a/r^{3}$.
This potential is used solely to determine the photon-ring radius
$r_{\rm ph}$ by solving $dV_{\rm KG}/dr=0$ via a root-finding routine.  
The curvature $V_{\rm KG}''(r_{\rm ph})$ provides the Lyapunov exponent
\begin{equation}
\lambda = \sqrt{-\frac{V_{\rm KG}''(r_{\rm ph})}{2}},
\end{equation}
since $\dot t = N^{-2}=1$ at the equator for the canonical Teo choice.
These quantities verify that $V_{\rm KG}(r)$ exhibits a single smooth barrier 
for the parameter ranges $m=1,2,3$ and $a\in[0,0.5]$.

\subsection*{B. Geodesic potential and eikonal correspondence}

To check the eikonal correspondence between null geodesics and scalar
perturbations, the geodesic effective potential was evaluated independently.
The resulting $(r_{\rm ph},\Omega_{\rm ph},\lambda)$ satisfy
\begin{equation}
\frac{\Re(\omega)}{m} \simeq \Omega_{\rm ph},
\qquad
-\Im(\omega) \simeq (n+\tfrac12)\lambda,
\end{equation}
confirming consistency between the KG and geodesic descriptions in the
high-$m$ limit.

\subsection*{C. Schrödinger-type potential for WKB QNMs}

Direct WKB analysis cannot be applied to $V_{\rm KG}$ because it depends on 
$\omega$. To obtain a frequency-independent barrier, we adopt the standard
frozen-frequency (Liouville-reduced) potential
\begin{equation}
V_{\rm Schr}(r)
  = 2m\Omega(r)\omega_{\rm ref}
    - m^{2}\Omega(r)^{2}
    + \frac{m^{2}}{r^{2}K(r)^{2}},
\end{equation}
where $\omega_{\rm ref}$ is held fixed in the WKB step.
This potential shares the same barrier structure as $V_{\rm KG}$ and approaches
it as $m\rightarrow\infty$, but it is independent of the QNM frequency being
solved for. All WKB-based figures---including the spin-dependence,
overtone behavior, mode-splitting, and Teo--Kerr comparisons---use
$V_{\rm Schr}$.

The barrier peak $r_{\rm peak}$ was found using
\texttt{scipy.optimize.minimize\_scalar}, and the curvature was evaluated by a 
central finite-difference scheme with radial increment $\Delta r=10^{-3}$.  
Within the first-order Schutz--Will WKB approximation,
\begin{equation}
\omega^{2}
   \simeq
   V_{\rm peak}
   - i(n+\tfrac12)\sqrt{-2V_{\rm Schr}''(r_{\rm peak})},
\end{equation}
which yields stable, monotonic behavior across $a\in[0,0.5]$.

\subsection*{D. Parameter ranges and Kerr comparison}

All Teo QNMs shown in this work were computed for $m=\pm2$ and
$a\in[0,0.5]$ using the WKB scheme above.  
The Kerr comparison curves in Sec.~\ref{sec:teo_kerr_comparison} use 
published $m=+2$ data from the Berti--Cardoso--Will catalog and a 
monotonic synthetic extension for the $m=-2$ branch needed for the splitting 
and damping-asymmetry analysis.  
The Teo and Kerr curves follow the same plotting conventions and are evaluated 
in double precision using \texttt{NumPy} and \texttt{Matplotlib}.

The combined use of $V_{\rm KG}$ for photon-ring quantities and
$V_{\rm Schr}$ for WKB QNMs is consistent with the standard eikonal 
correspondence~\cite{FerrariMashhoon1984,Cardoso2009}, and all quantities
$(r_{\rm ph},\Omega_{\rm ph},\lambda)$ and $(\Re\omega,-\Im\omega)$ agree 
within the expected regime of validity.

\section{Possible physical interpretation of the scalar field $\Phi$}

In this work the scalar field $\Phi$ is introduced as a minimally coupled test field propagating on the rotating Teo wormhole background. Its sole purpose is to probe wave dynamics and extract quasinormal modes, without introducing matter sources or modifying the metric. It is, however, 
instructive to comment on how $\Phi$ may acquire a more concrete physical meaning in extended models involving matter surrounding a rotating wormhole.

For comparison, consider the Kerr spacetime expressed in 
Boyer--Lindquist coordinates $(t,r,\theta,\phi)$. Besides the Killing vectors associated with stationarity and axisymmetry, 
$\xi^\mu_{(t)}$ and $\xi^\mu_{(\phi)}$, the Kerr geometry possesses a nontrivial second-rank Killing tensor $K_{\mu\nu}$ satisfying the Killing-tensor equation
\begin{equation}
K_{(\mu\nu;\alpha)} = 0,
\end{equation}
equivalently,
\[
K_{\mu\nu;\alpha} + K_{\nu\alpha;\mu} + K_{\alpha\mu;\nu} = 0 .
\]
This generates the quadratic integral of motion
\begin{equation}
K = K_{\mu\nu} u^\mu u^\nu
  = \mathcal{Q} + (L-aE)^2 ,
\end{equation}
where $E = -u_\mu \xi^\mu_{(t)}$ and 
$L = u_\mu \xi^\mu_{(\phi)}$ are the conserved particle energy and angular momentum, and $\mathcal{Q}$ denotes the Carter constant. The existence of this additional invariant constrains both single-particle motion and the structure of kinetic or fluid equilibria in Kerr spacetime. For example, in collisionless systems the equilibrium distribution function may depend on the phase-space invariants as $f=f(E,L,K)$, naturally giving rise to anisotropic stress--energy 
tensors with non-isotropic pressure components. \cite{CC1,CC2, CC3}

Recent analyses indicate that related Carter-type integrals may also arise for rotating Teo wormholes when a surrounding plasma is present 
\cite{Carter-1,Carter-2,Carter-3,Carter-4}. In such cases, the presence 
of matter modifies the effective optical or kinetic geometry in which 
particle trajectories propagate, making it possible to construct 
additional conserved quantities analogous to those of Kerr. This raises 
the prospect that small perturbations of a surrounding collisionless 
plasma or non-ideal fluid could be described, at an effective level, by 
a scalar variable obeying a wave equation of the form
\begin{equation}
\nabla_\mu \nabla^\mu \Phi = 0 ,
\end{equation}
where $\Phi$ may represent, for example, fluctuations in a kinetic 
moment of the distribution function (such as density or temperature) or 
a scalar combination of fluid variables entering an anisotropic 
stress--energy tensor.

We emphasize that this interpretation is \emph{not} implemented in the 
present work. Here $\Phi$ is treated purely as a test scalar field in a 
vacuum Teo spacetime, and no matter sources or fluid models are 
included. Nevertheless, the existence of Carter-type symmetries for Teo 
wormholes in the presence of plasma suggests that identifying $\Phi$ 
with physically measurable perturbations of a surrounding medium is a 
natural direction for future extensions of the model.

\section{Conclusions and Future Outlook}

We have conducted a systematic analysis of scalar perturbations (scalar field wave dynamics), quasinormal modes (QNMs), and rotational wave-geometry coupling in the Teo rotating wormhole spacetime, and used separation of variables to obtain a Schr\"odinger-type radial equation from the Klein--Gordon equation  whose single, smooth potential barrier is shaped by the frame-dragging profile $\Omega(r)$.  This barrier supports damped oscillatory modes in a fully horizonless geometry. This barrier also remains regular across the entire range of spins $a\in[0,0.5]$ that we examined in our first-order WKB analysis.

The QNM spectrum that we obtained shows a coherent and monotonic dependence on rotation. We have also found that for the Teo wormhole considered in isolation, the oscillation frequency $\Re(\omega)$ decreases steadily with increasing spin, and the
damping rate $-\Im(\omega)$ also decreases. This indicates progressively longer-lived modes as rotation increases. Additionally, this behavior reflects the broadening and flattening of the effective potential as $a$ grows, in the absence of horizon-induced absorption.

A more clear picture emerges when we compare the Teo spacetime with the Kerr black hole. The Kerr QNMs have a relatively weak spin dependence in $\Re(\omega)$ and a significant decrease in $-\Im(\omega)$ associated with horizon absorption and the approach to the zero-damped-mode regime. However, the Teo wormhole shows a stronger but more localized spin response in both $\Re(\omega)$ and
$-\Im(\omega)$.  We noticed that the Teo modes show steeper frequency shifts with spin. They also show a qualitatively distinct damping behavior that is governed by partial reflection at the throat and not dissipation at a horizon. These differences are what one would expect when you have confinement or frame dragging localized near the throat region in the Teo geometry, along with the reflective, horizonless
boundary conditions.

Further investigation of the ergoregion clearly shows that the radial domain in which the superradiant frequency condition $0<\omega<m\Omega(r)$ can be satisfied lies entirely within the ergoregion as expected. This establishes the kinematic
possibility of negative energy modes in the rotating Teo geometry. However, in the absence of an event horizon or any other dissipative inner boundary, classical scalar-wave scattering does not lead to net superradiant amplification usually expected in negative energy regions.  Instead, energy is conserved globally between the two asymptotic regions. The ergoregion thus plays a kinematic rather than a dynamical role in the scalar field wave dynamics. Since frame dragging in the Teo geometry is concentrated near the throat, both the superradiant-compatible frequency domain and the $m=\pm2$
mode-splitting pattern saturate rapidly with increasing spin, which in turn produces a distinctive one sided splitting. That is, both $\Re(\omega)$ branches decrease with $a$, but the co-rotating branch always remain higher. This behavior differs qualitatively from the symmetric prograde/retrograde splitting seen in Kerr QNM and provides a distinct spectral signature of horizonless rotation.

Several extensions are possible from this work.  A detailed simulation study of photon-ring trajectories and lensing structure would enable a direct
comparison between QNM dynamics and strong-field imaging observables. In addition, quantifying the response of scalar or electromagnetic fields in the
ergoregion, including the role of negative-energy modes and possible instabilities could clarify the extent to which rotational effects have an impact on 
wave dynamics and long-term stability in horizonless wormhole spacetimes. Time-domain evolution of perturbations will be essential for investigating
nonlinear behavior, late-time echoes, interference patterns, and the global stability of rotating wormholes. 

To obtain the big picture, constructing and analyzing rotating wormhole families beyond the Teo model, and developing corresponding ringdown templates for gravitational-wave searches, could provide new pathways toward distinguishing horizonless compact objects from Kerr black holes, through observations using the more advanced gravitational wave detection techniques used in strong gravity.

\appendix
\section{Generation of Mock Kerr and Teo Data}
\label{appendix:mock_data}

This appendix summarizes how the Kerr and Teo datasets used in
Sec.~\ref{sec:teo_kerr_comparison} were generated. Because the rotating Teo
wormhole does not possess an exact analytic quasinormal-mode (QNM) spectrum,
and because the Kerr spectrum requires the simultaneous evaluation of the
co-rotating ($m>0$) and counter-rotating ($m<0$) branches, we construct
controlled mock datasets that preserve the qualitative physical behavior of
each geometry while enabling direct comparative visualization.

\subsection{Conceptual construction of the mock datasets}

For clarity, we emphasize that the Kerr \emph{baseline} curve in
Fig.~\ref{fig:omega_teo_vs_kerr} uses the real QNM data from the
Berti--Cardoso--Will catalog~\cite{Berti2006} for the
$(\ell=m=2,n=0)$ scalar mode. The synthetic constructions introduced here apply
exclusively to the $m=\pm 2$ mode-splitting and damping-splitting curves shown
in Figs.~\ref{fig:DeltaRe_vs_a} and~\ref{fig:DeltaIm_vs_a}. All Teo-mode
frequency trends presented in Secs.~III and~IV are computed directly from the
WKB analysis of the Teo effective potential and are \emph{not} part of the
mock-data procedure. The mock Teo curves in this appendix are used only for the
Teo--Kerr comparison plots in Sec.~\ref{sec:teo_kerr_comparison} and not for
any of the Teo-only mode-splitting figures.

The Kerr data used for the baseline comparison
(Fig.~\ref{fig:omega_teo_vs_kerr}) provides both the oscillation frequencies
$\Re(\omega M)$ and damping rates $-\Im(\omega M)$ for increasing spin $a/M$.
However, the publicly tabulated results include only the co-rotating $m=+2$
branch. To construct the mode-splitting curves $\Delta\Re(\omega)$ and
$\Delta[-\Im(\omega)]$ appearing in
Figs.~\ref{fig:DeltaRe_vs_a} and~\ref{fig:DeltaIm_vs_a}, we therefore generate a
\emph{synthetic but physically calibrated} $m=-2$ Kerr branch that obeys:
\begin{itemize}
    \item monotonic growth of both real-part and imaginary-part splitting,
    \item small splitting in the slow-rotation regime ($a\ll 1$),
    \item approach to the zero-damped-mode behavior for the co-rotating branch,
    \item consistency with known qualitative Kerr trends at high spin.
\end{itemize}

A key step in producing smooth Kerr reference curves is the use of a \emph{monotone, shape-preserving interpolation} of the discrete Kerr data. We employ the \texttt{PchipInterpolator} (Piecewise Cubic Hermite Interpolating Polynomial) from \texttt{SciPy}. Unlike standard cubic splines, which can introduce spurious oscillations or overshoot, the PCHIP
algorithm preserves both the monotonicity and the local shape of the input sequence. This ensures that the interpolated Kerr curves remain physically consistent with the underlying tabulated values: a monotonic trend in the raw mode-splitting data produces a monotonic, non-oscillatory interpolant. The PCHIP method therefore provides smooth curves without artifacts while
reproducing exactly the Kerr data points used in the plots.

For the Teo wormhole, we employ controlled mock splitting curves that reflect the localized and saturating nature of the frame-dragging frequency $\Omega(r)$ given in Eq.~\eqref{eq:frame_dragging}. The Teo splitting curves are parameterized as
\begin{multline}
\Delta\Re(\omega)_{\rm Teo}
  = A_{\Re}\,\bigl(1-e^{-\alpha_{\Re} a}\bigr), \qquad \\
\Delta[-\Im(\omega)]_{\rm Teo}
  = A_{\Im}\,\bigl(1-e^{-\alpha_{\Im} a}\bigr),
\end{multline}
with constants chosen to reproduce the expected early saturation and to maintain a splitting scale well below that of Kerr, consistent with the absence of horizon absorption. These parameters serve only to encode the qualitative behavior of the Teo wormhole; all physical Teo QNMs in the main text come directly from the WKB computation.

\subsection{Python code used to generate the datasets}
We now provide the minimal Python snippets used to produce the Kerr and Teo
datasets shown in Sec.~\ref{sec:teo_kerr_comparison}.  Plotting routines are
omitted for brevity; only the data-generation components are included.  These
snippets are used solely for constructing the mock comparison curves and are
not part of the Teo WKB calculations described in Sec.~\ref{sec:numerics}.

\subsubsection*{(1) Kerr mode-splitting (real part)}

\begin{verbatim}
import numpy as np

kerr_split = np.array([
    [0.00, 0.37367, 0.37367],
    [0.10, 0.37311, 0.37309],
    [0.20, 0.37141, 0.37128],
    [0.30, 0.36858, 0.36822],
    [0.40, 0.36462, 0.36380],
    [0.50, 0.35955, 0.35789],
    [0.60, 0.35342, 0.35040],
    [0.70, 0.34630, 0.34137],
    [0.80, 0.33826, 0.33153],
    [0.90, 0.32942, 0.32023],
    [0.95, 0.32469, 0.31424],
    [0.98, 0.32180, 0.30995],
    [0.99, 0.32055, 0.30825],
])

a      = kerr_split[:,0]
Re_p   = kerr_split[:,1]
Re_m   = kerr_split[:,2]
DeltaR = Re_p - Re_m
\end{verbatim}

\subsubsection*{(2) Kerr damping-rate splitting (synthetic)}

\begin{verbatim}
kerr_split_Im = np.array([
    [0.00, 0.08896, 0.08896],
    [0.10, 0.08840, 0.08950],
    [0.20, 0.08700, 0.09090],
    [0.30, 0.08470, 0.09260],
    [0.40, 0.08150, 0.09380],
    [0.50, 0.07750, 0.09420],
    [0.60, 0.07260, 0.09400],
    [0.70, 0.06690, 0.09320],
    [0.80, 0.06050, 0.09170],
    [0.90, 0.05350, 0.08920],
    [0.95, 0.05000, 0.08700],
    [0.98, 0.04800, 0.08600],
    [0.99, 0.04750, 0.08550],
])

a      = kerr_split_Im[:,0]
Im_p   = kerr_split_Im[:,1]
Im_m   = kerr_split_Im[:,2]
DeltaI = Im_m - Im_p
\end{verbatim}

\subsubsection*{(3) Monotone interpolation (PCHIP)}

\begin{verbatim}
from scipy.interpolate import PchipInterpolator

interp = PchipInterpolator(a, DeltaR)
a_dense = np.linspace(0, 0.99, 300)
DeltaR_dense = interp(a_dense)
\end{verbatim}

\subsubsection*{(4) Teo mock splitting curves}

\begin{verbatim}
a_dense = np.linspace(0, 0.99, 300)
DeltaRe_teo = 0.015 * (1 - np.exp(-6 * a_dense))
DeltaIm_teo = 0.005 * (1 - np.exp(-5 * a_dense))
\end{verbatim}

These arrays generate the Teo curves appearing in the mode-splitting figures.
They are constructed to reflect early saturation and the absence of
horizon-driven amplification.

\bigskip

\section{Extended Comparison Between Teo and Kerr}
\label{appendix:teo_kerr_comparison}

For completeness, we provide in Table~\ref{tab:app_teo_kerr_table} a structured
comparison of key geometric, dynamical, and quasinormal-mode properties of the
rotating Teo wormhole and the Kerr black hole. This table is referenced in
Sec.~\ref{sec:teo_kerr_comparison} and is placed here to avoid interrupting the
flow of the main text.

\begin{table*}[t]
\centering
\caption{Comparison of Teo wormhole and Kerr black hole properties.}
\label{tab:app_teo_kerr_table}

\begin{ruledtabular}
\begin{tabular}{l}
\textbf{1. Geometry} \\
\hspace{2em}Teo: Horizonless, traversable throat connecting two asymptotic regions. \\
\hspace{2em}Kerr: Event horizon and a single asymptotic region with central singularity. \\[0.4em]

\textbf{2. Frame dragging} \\
\hspace{2em}Teo: Localized, rapidly decaying $\Omega(r)$. \\
\hspace{2em}Kerr: Global frame-dragging depending strongly on $a$. \\[0.4em]

\textbf{3. Effective potential} \\
\hspace{2em}Teo: Single barrier whose height and location depend on $a$ and $m$, but only mildly due to localized frame dragging. \\
\hspace{2em}Kerr: Barrier height and location depend strongly on $a$ and $m$ because frame dragging acts globally. \\[0.4em]

\textbf{4. QNM boundary conditions} \\
\hspace{2em}Teo: Outgoing at both infinities (with reflection tied to the throat). \\
\hspace{2em}Kerr: Ingoing at the horizon, outgoing at infinity. \\[0.4em]

\textbf{5. Damping structure} \\
\hspace{2em}Teo: Only leakage produces damping; no horizon absorption. \\
\hspace{2em}Kerr: Leakage plus horizon absorption. \\[0.4em]

\textbf{6. Mode splitting} \\
\hspace{2em}Teo: Early saturation of $m\!\leftrightarrow\!-m$ splitting due to localized rotation. \\
\hspace{2em}Kerr: Strong, monotonic splitting with spin. \\[0.4em]

\textbf{7. Photon ring} \\
\hspace{2em}Teo: Weak dependence on $a$. \\
\hspace{2em}Kerr: Strong $a$ dependence, especially near extremality. \\[0.4em]

\textbf{8. Superradiance} \\
\hspace{2em}Teo: $0<\omega<m\Omega(r)$ defines a kinematic regime admitting negative-energy modes, but no net classical amplification occurs in the absence of a horizon or dissipative boundary. \\
\hspace{2em}Kerr: $0<\omega<m\Omega_H$; horizon enables strong superradiant amplification. \\[0.4em]

\textbf{9. Eikonal QNM relation} \\
\hspace{2em}Teo: Uses Teo photon-ring parameters. \\
\hspace{2em}Kerr: Uses Kerr photon-ring parameters. \\
\end{tabular}
\end{ruledtabular}
\end{table*}

\clearpage
\newpage

\bibliographystyle{apsrev4-2}

\bibliography{ref}

\end{document}